\documentclass[pra,10pt,nofootinbib, onecolumn,notitlepage, superscriptaddress]{revtex4-1}

\pdfoutput=1

\usepackage{amsmath}
\usepackage{amsfonts}
\usepackage{braket}
\usepackage{amssymb}
\usepackage{mathbbol}
\usepackage{graphicx}
\usepackage[T1]{fontenc} 
\usepackage{slashed}
\graphicspath{{.}{figures/}} 

\usepackage[colorlinks=true,linkcolor=blue, citecolor=blue, urlcolor=blue, bookmarks]{hyperref}

\newcommand{\vect}[1]{\boldsymbol{#1}}

\newcommand{\sh}[1]{\slashed{#1}}
\newcommand{\da}{\downarrow}
\newcommand{\ua}{\uparrow}

\begin{document}
	
	\title{The light front wave functions and diffractive electroproduction of vector mesons.}
	
\author{Chao Shi}
\email[]{cshi@nuaa.edu.cn}
\affiliation{Department of Nuclear Science and Technology, Nanjing University of Aeronautics and Astronautics, Nanjing 210016, China}

\author{Ya-Ping Xie}
\email[]{xieyaping@impcas.ac.cn}
\affiliation{Institute of Modern Physics, Chinese Academy of Sciences, Lanzhou 730000, China}

\author{Ming Li}
\affiliation{Department of Nuclear Science and Technology, Nanjing University of Aeronautics and Astronautics, Nanjing 210016, China}

\author{Xurong Chen}
\affiliation{Institute of Modern Physics, Chinese Academy of Sciences, Lanzhou 730000, China}
\affiliation{Guangdong Provincial Key Laboratory of Nuclear Science, Institute of Quantum Matter, South China Normal University, Guangzhou 510006, China}

\author{Hong-Shi Zong}
\affiliation{Department of Physics, Nanjing University, Nanjing 210093, China}

\begin{abstract}
We determine the leading Fock-state light front wave functions (LF-LFWFs) of the $\rho$ and J/$\psi$ mesons, for the first time from the Dyson-Schwinger and Bethe-Salpeter equations (DS-BSEs) approach. A unique advantage of this method is that it renders a direct extraction of LF-LFWFs in presence of a number of higher Fock-states.  Modulated by the current quark mass and driven by the dynamical chiral symmetry breaking (DCSB), we find the $\rho$ and $J/\psi$ LF-LFWFs different in profile, i.e., the former are broadly distributed in $x$ (the longitudinal light-cone momentum fraction of meson carried by quark) while the latter are narrow. Moreover, the $\rho$ LF-LFWFs contribute less than 50\% to the total Fock-state normalization, suggesting considerable higher Fock-states in $\rho$.  We then use these LF-LFWFs to study the diffractive $\rho$ and $J/\psi$ electroproduction  within the dipole picture. The calculated cross section shows general agreement with HEAR data, except for growing discrepancy in $\rho$ production at low photon virtuality. Our work provides a first dipole picture analysis on diffractive $\rho$ electroproduction that confronts the parton nature of the light (anti)quarks. 
	\end{abstract}
	
	\maketitle

	
\section{Introduction}
\label{sec:intro}
Diffractive vector meson production provides an important probe to the gluon saturation at small $x$ \cite{Ryskin:1992ui}. Within the dipole picture, saturation and non-saturation scattering amplitudes yield sizable effects in, e.g., the t-distribution of differential cross section \cite{Armesto:2014sma}, the cross section ratio between eAu$\rightarrow$eAuV and ep$\rightarrow$epV \cite{Accardi:2012qut}. Meanwhile, the LF-LFWFs of vector mesons and photon are important nonperturbative element of the dipole picture. Their determination in connection with QCD greatly helps reduce the theoretical uncertainties, and substantially deepens our understanding of the hard diffractions.

While the non-relativistic QCD (NRQCD) sheds light on heavy meson LFWFs \cite{Ryskin:1992ui,Brodsky:1994kf,Lappi:2020ufv}, it remains a great challenge to calculate light vector meson LFWFs in connection with QCD to date. The light-cone QCD Hamiltonian, which encodes abundant creation and annihilation of light-quarks and gluons, gets intensely difficult to diagonalize with increasing number of Fock-states \cite{Brodsky:1997de}. Therefore, existing dipole picture studies all employ phenomenological (or effective) $\rho$ LF-LFWFs within the constituent quark picture \cite{Kowalski:2003hm,Forshaw:2010py,Forshaw:2012im,Ahmady:2016ujw}, i.e, it admits an effective quark mass $m_{u/d}=[46, 140]$ MeV and  excludes (or  effectively absorb) the higher Fock-states. However, their relation to the parton nature of light (anti)quarks in real QCD remains elusive.

Regarding the large uncertainties within vector meson LFWFs (particularly the light ones such as $\rho$ and/or $\phi$), we here tackle this problem with a novel approach based on DS-BSEs study \cite{tHooft:1974pnl,Liu:1992dg,Burkardt:2002uc}. The modern DS-BSEs study is closely connected with QCD, i.e., it incorporates the quark and gluon degrees of freedom and selectively re-sums infinitely many Feynman diagrams while respecting various symmetries of QCD \cite{Roberts:1994dr,Bashir:2012fs}, i.e., prominently the poincare symmetry and chiral symmetry. We then project the $\rho$ and $J/\psi$ covariant BS wave functions onto the light front and extract the LF-LFWFs from the many Fock-states embedded \cite{Mezrag:2016hnp,Shi:2018zqd,dePaula:2020qna}. As will be shown, these LFWFs directly characterize the parton structure of vector mesons. With them, the  dipole approach will be confronted with diffractive $\rho$ electroproduction at HERA within the parton picture for the first time.

\section{Leading Fock state light front wave functions of $\rho$ and $J/\psi$:}  
\subsection{Classification of vector meson LF-LFWFs}
\label{sec:intro}
The leading Fock-state configuration of a vector meson state  is expressed with the nonperturbative LFWFs $\Phi^\Lambda_{\lambda,\lambda'}$
\begin{align}\label{eq:LFWF1}
|M\rangle^\Lambda &= \sum_{\lambda,\lambda'}\int \frac{d^2 \vect{k}_T}{(2\pi)^3}\,\frac{dx}{2\sqrt{x\bar{x}}}\, \frac{\delta_{ij}}{\sqrt{3}} \nonumber \\
&\hspace{10mm} \Phi^\Lambda_{\lambda,\lambda'}(x,\vect{k}_T)\, b^\dagger_{f,\lambda,i}(x,\vect{k}_T)\, d_{f,\lambda',j}^\dagger(\bar{x},\bar{\vect{k}}_T)|0\rangle.
\end{align}
Here the $\vect{k}_T=(k^x,k^y)$ is the transverse momentum of the  quark $f$, and $\bar{\vect{k}}_T=-\vect{k}_T$ for antiquark $\bar{f}$. The longitudinal momentum fraction carried by quark is $x=\frac{k^+}{P^+}$, and $\bar{x}=1-x$ for antiquark. The $i$ and $j$ are color indices. The quark helicity $\lambda$ runs through  $\ua$ and $\da$, while the meson helicity $\Lambda$ runs through $0$ and $\pm 1$. The $\Phi(x,\vect{k}_T)$'s can be further expressed with amplitudes $\psi(x,\vect{k}_T^2)$'s which contain only scalars arguments $x$ and $\vect{k}_T^2$ \cite{Ji:2003fw}. Denoting the quark helicity $\ua=+$ and $\da=-$, and omitting the function arguments, one finds for longitudinally polarized mesons
\begin{align}\label{eq:phiL}
\hspace{00mm}\Phi_{\pm,\mp}^{0}&=\psi^{0}_{(1)},  \ \ \ \ \ 
\Phi_{\pm,\pm}^{0}=\pm k_T^{(\mp)} \psi^{0}_{(2)},
\end{align}
with $k_T^{(\pm)}=k^x \pm i k^y$, and for transversely polarized  mesons ($\Lambda=\pm 1$)
\begin{align}\label{eq:phiT}
\Phi_{\pm,\pm}^{\pm 1}&=\psi^{1}_{(1)},
&\Phi_{\pm,\mp}^{\pm 1}&=\pm  k_T^{(\pm)}\psi^{1}_{(2)}, \notag \\
\Phi_{\mp,\pm}^{\pm 1}&=\pm k_T^{(\pm)}\psi^{1}_{(3)},
&\Phi_{\mp,\mp}^{\pm 1}&=(k_T^{(\pm)})^2\psi^{1}_{(4)}.
\end{align}
Note the $\Lambda=-1$ meson can be  obtained from $\Lambda=+1$  with a $\hat{Y}$ transform, which consists a parity operation followed by a 180\textdegree \  rotation around the y axis \cite{Ji:2003yj}. With the help of $\hat{Y}$ and charge parity, we find the constraints
\begin{align}\label{eq:psi1}
\psi_{(i)}^{\Lambda}(x,\vect{k}_T^2)=\psi_{(i)}^{\Lambda}(1-x,\vect{k}_T^2),
\end{align}
with one exception 
\begin{align}\label{eq:psi2}
\psi^{1}_{(2)}(x,\vect{k}_T^2)&=-\psi^{1}_{(3)}(1-x,\vect{k}_T^2).
\end{align}
In the end, for $\rho^0$ (with isospin symmetry) or $J/\psi$, there are totally \emph{five} independent $\psi^\Lambda_{(i)}$'s at leading Fock-state.

\subsection{From vector meson Bethe-Salpeter wave functions to LF-LFWFs.}

Within the DS-BSEs framework, the vector mesons can be solved with their covariant Bethe-Salpter wave functions. In practice, this is achieved by taking the rainbow-ladder (RL) truncation and aligning the quark's DSE for full quark propagator $S(p)$ and meson's BSE  for BS wave functions $\Gamma_\mu^M(k,P)$ \cite{Maris:1999nt}, i.e.,
\begin{eqnarray}
S(p)^{-1} &=& Z_2 \,(i\gamma\cdot p + m^{\rm bm}) + Z_2^2 \int^\Lambda_\ell\!\! {\cal G}(\ell)
\ell^2 D_{\mu\nu}^{\rm free}(\ell)
\frac{\lambda^a}{2}\gamma_\mu S(p-\ell) \frac{\lambda^a}{2}\gamma_\nu ,
\label{rainbowdse} \rule{1em}{0ex}\\
\Gamma^M_\mu(k;P) &=&  - Z_2^2\int_q^\Lambda\!\!
{\cal G}((k-q)^2)\, (k-q)^2 \, D_{\mu\nu}^{\rm free}(k-q)
\frac{\lambda^a}{2}\gamma_\mu S(q_+)\Gamma^M_\mu(q;P) S(q_-) \frac{\lambda^a}{2}\gamma_\nu ,
\label{ladderBSE}
\end{eqnarray}
Here $\int^\Lambda_q$ implements a Poincar\'e invariant regularization of the four-dimensional integral, with $\Lambda$ the regularization mass-scale. $D^{\rm{free}}_{\mu\nu}$ is the free gluon propagator. $m^{\rm bm}(\Lambda)$ is the current-quark bare mass. $Z_{2}$ is the quark wave function renormalisation constants at renormalization point $\mu$. Here a factor of $1/Z_2^2$ is picked out to preserve multiplicative renormalizability in solutions of the gap and Bethe-Salpeter equations \cite{Bloch:2002eq}. The Bethe-Salpeter amplitudes are eventually normalized canonically (see, e.g., Eq.\,(25) in Ref.~\cite{Maris:1999nt}).

The modeling function ${\cal G}(l^2)$ absorbs the strong coupling constant $\alpha_s$, as well as dressing effect from quark-gluon vertex and full gluon propagator. Popular models include the earlier Maris-Tandy (MT) model, and the later Qin-Chang (QC) model
\cite{Qin:2011xq}
\begin{equation}
\label{eq:GQC}
{\cal G}_{QC}(s) = \frac{8 \pi^2}{\omega^4} D  \, {\rm e}^{-s/\omega^2}
+ \frac{8 \pi^2 \gamma_m}{\ln [ \tau + (1+s/\Lambda_{\rm QCD}^2)^2]} {\cal F}(s).
\end{equation}
The first term models the infrared behavior, and the second term is perturbative QCD result \cite{Maris:1997tm,Qin:2011xq}.  The QC model improves the infrared part of MT model to be in concert with modern gauge sector study, while in hadron study the two are equally good.  Combined with the RL truncation, the MT and/or QC model  well describes a range of hadron properties, including the pion and $\rho$ meson masses, decay constants and various elastic and transition form factors \cite{Maris:1997hd,Maris:1999nt,Maris:2000sk,Jarecke:2002xd,Bhagwat:2006pu,Xu:2019ilh}. The success also extends to nucleon by solving the Faddeev equation \cite{Eichmann:2009qa,Eichmann:2011vu}. These achievements owe greatly to the nice property of the RL truncation by preserving the (near) chiral symmetry of QCD (respecting the axial vector Ward-Takahashi identity) \cite{Maris:1997hd}. It is therefore  capable of simultaneously describing the almost massless pion as a Goldstone boson and the much more massive $\rho$ and nucleon, reflecting different aspects of the DCSB. Here we will explore the prediction of RL DS-BSEs on the vector meson LF-LFWFs.

Having solved Eqs.~(\ref{rainbowdse}-\ref{ladderBSE}) and obtain $S(p)$ and $\Gamma_\mu^M(q;P)$, we then project the  vector meson BS wave function $\chi^M_\mu(k,P) = S(k+P/2)\,\Gamma_\mu^M(k,P)\,S(k-P/2)$ onto the light front to obtain the LF-LFWFs using
\begin{align}\label{eq:chi2phi}
\Phi^\Lambda_{\lambda,\lambda'}(x,\vect{k}_T)&=-\frac{1}{2\sqrt{3}}\int \frac{dk^- dk^+}{2 \pi} \delta(x P^+-k^+) \textrm{Tr}\left [\Gamma_{\lambda,\lambda'}\gamma^+ \chi^M(k,P) \cdot \epsilon_\Lambda(P) \right ].
\end{align}
This can be derived by generalizing the projection method for pseudo-scalar meson in \cite{Liu:1992dg,Burkardt:2002uc}. Here the $\epsilon_\Lambda(P)$ is the  meson polarization vector. The $\Gamma_{\pm,\mp}=I\pm \gamma_5$ and $\Gamma_{\pm,\pm}=\mp(\gamma^1\mp i\gamma^2)$ projects out certain (anti)quark helicity configuration.
The $\chi^M_\mu(k,P)$ can be expressed  with the dressed quark propagator $S(k)$  and  BS amplitude $\Gamma^M_\mu(k,P)$ as . The trace is taken over Dirac, color and flavor indices. An implicit color factor $\delta_{ij}$ is associated with $\Gamma_{\lambda,\lambda'}$, as well as a flavor factor diag$(1/\sqrt{2},-1/\sqrt{2})$ for $\rho^0$. Then we calculate the $(2x-1)$-moments of $\psi^\Lambda_{(i)}(x,\vect{k}_T^2)$ (which are equivalent to $\Phi_{\lambda,\lambda'}^\Lambda$ through Eqs.~(\ref{eq:phiL}-\ref{eq:phiT})) at every $|\vect{k}_T|$, i.e.,
\begin{align}\label{eq:mom}
\langle (2x-1)^m \rangle_{|\vect{k}_T|}^{(i)}&=\int dx (2x-1)^m \psi^\Lambda_{(i)}(x,\vect{k}_T^2),
\end{align}
with $m=0,1,2,..$. From these  moments we reconstruct the LFWFs. For practical reasons, we treat the $\rho$ and $J/\psi$ with somewhat different techniques.

In solving the $\rho$ DS-BSEs, we only take the infrared part of the QC model, i.e., the first term on the right hand side of Eq.~(\ref{eq:GQC}). We refer to it as QC-IR model. Since the support of light quark propagator and BS amplitude are dominated by low relative momentum, the ultraviolet term of QC model has relatively small effect. Such treatment was also employed in other light quark sector studies \cite{Fischer:2009jm,Chang:2009zb}. Adopting the well-determined parameters $\omega=0.5\,$GeV, $D = (0.82\,{\rm GeV})^3/\omega$ \cite{Qin:2011xq,Shi:2014uwa,Xu:2019ilh} and the current quark mass $m_{u/d}=5$ MeV, we reproduce $m_\pi=131$ MeV and $f_\pi=90$ MeV, as well as $m_\rho=717$ MeV and $f_\rho=140$ MeV comparing to experimental values  $m_\rho=775$ MeV and $f_\rho=156$ MeV \cite{Zyla:2020zbs}. We choose QC-IR model rather than QC model as it renders an exponentially $k^2-$suppressed $\Gamma^M_\mu(k,P)$.  This allows us to directly compute up to ninth-moment with Eq.~(\ref{eq:mom}), with the numerical noises heavily suppressed. Note with QC model, only the first two or three moments can be directly computed for now. We then fit the moments with a flexible parameterization \cite{Shi:2014uwa,Shi:2015esa}
\begin{align}
\label{eq:fit}
\psi_{(i)}^\Lambda(x,\vect{k}_T^2)&\approx [x (1-x)]^{\alpha-1/2}\!\!\!
\sum_{j=0,2}^{\ } a_j^{\alpha} C_j^{\alpha}(2 x -1)+ [x (1-x)]^{\alpha'-1/2}\,
\sum_{j'=1,3}^{ } a_{j'}^{\alpha'} C_{j'}^{\alpha'}(2 x -1),
\end{align}
where the $\alpha$, $\alpha'$, $a_j^\alpha$ and $a_{j'}^{\alpha'}$ are fitting parameters. They implicitly depend  on the $\vect{k}_T^2$, $\Lambda$ and $i$. 
The $C_j^\alpha(x)$ is the Gegenbaur polynomial of order $\alpha$, so the first term on the right hand side of Eq.~(\ref{eq:fit}) is symmetric in $x$ with respect to $x=1/2$, and the second term is anti-symmetric. They are devised to fit the even and odd (2x-1)-moments separately. Using Eq.~(\ref{eq:fit}), we well reproduce all the moments with deviations less than 1\%.

For $J/\psi$, we choose the parameters $\omega=0.7$ GeV and $D=(0.6\  \textrm{GeV})^3/\omega$ from a recent global analysis on heavy meson spectrum involving both charm and bottom quarks \cite{Chen:2020ecu}. They are a bit different from that in the light sector, as the the DCSB dressing effects they mimic are quantitatively different between light and heavy sectors. In principle, this deviation can be reduced by going beyond RL truncation. Meanwhile we keep the ultraviolet term of the QC model as it is more relevant for heavy quarks. With running quark mass $m_c(\mu=m_c)=1.33$ GeV, we get $m_{J/\psi}=3.09$ GeV and $f_{J/\psi}=300$ MeV, as compared to PDG data   $m_c(\mu=m_c)=1.28$ GeV, $m_{J/\psi}=3.096$ GeV  and $f_{J/\psi}=294$ MeV by leptonic decay $\Gamma(J/\psi\rightarrow e^+ e^-)=5.53$ keV \cite{Zyla:2020zbs}. To compute the $J/\psi$ LFWFs, we adopt the technique used in \cite{Shi:2018zqd,Shi:2020pqe}: by fitting the meson BS amplitude with Nakanishi-like representation \cite{Nakanishi:1963zz} and the quark propagator with pairs of complex conjugate poles form which is particularly accurate in heavy sector \cite{Souchlas:2010boa}, we are able to compute point-wisely accurate LFWFs. More details can be found in the appendix. 

\begin{figure}[!t]
	\centering
	\includegraphics[width=3.5in]{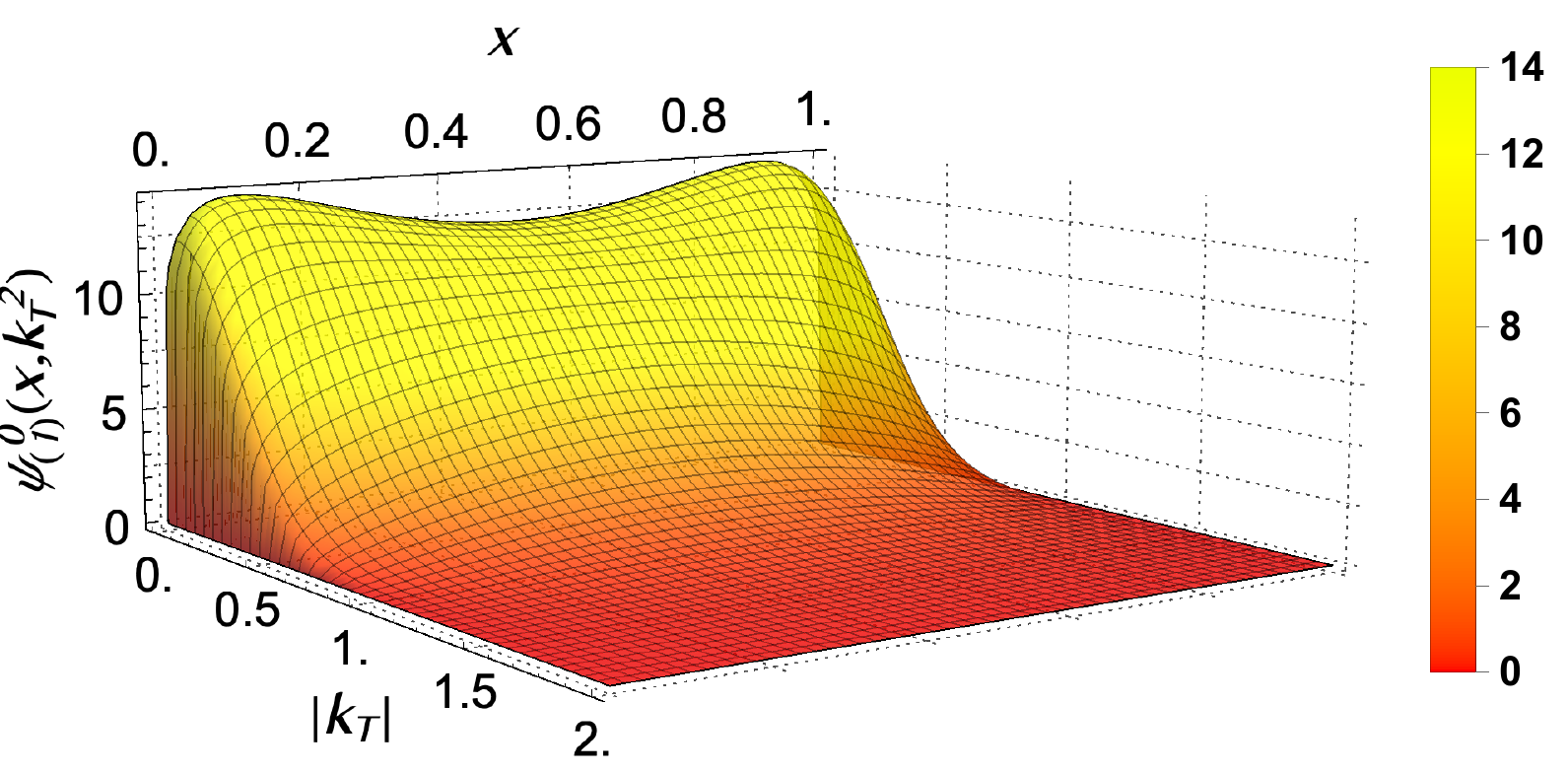}
	\includegraphics[width=3.5in]{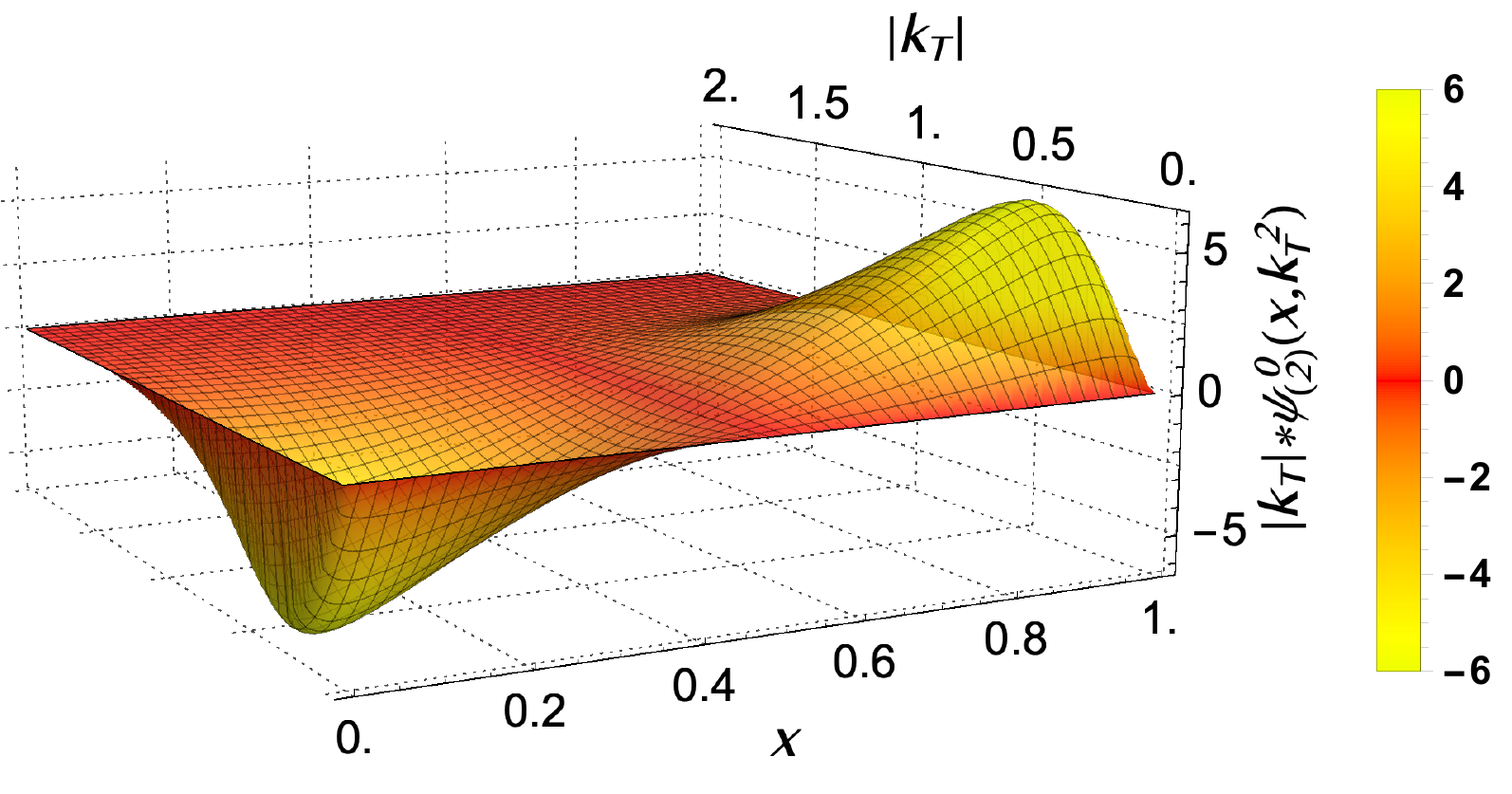}
		\includegraphics[width=3.5in]{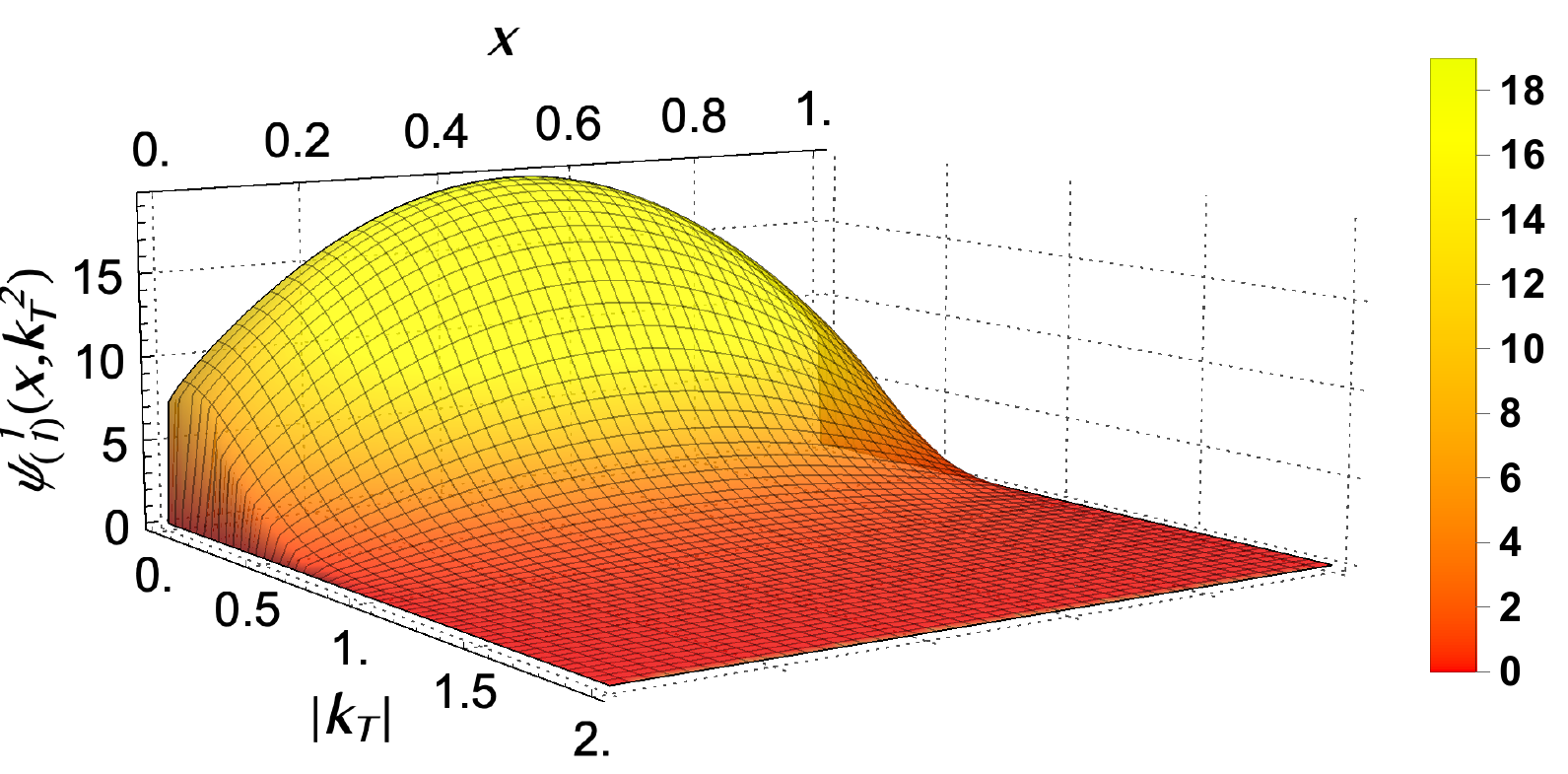}
	\includegraphics[width=3.5in]{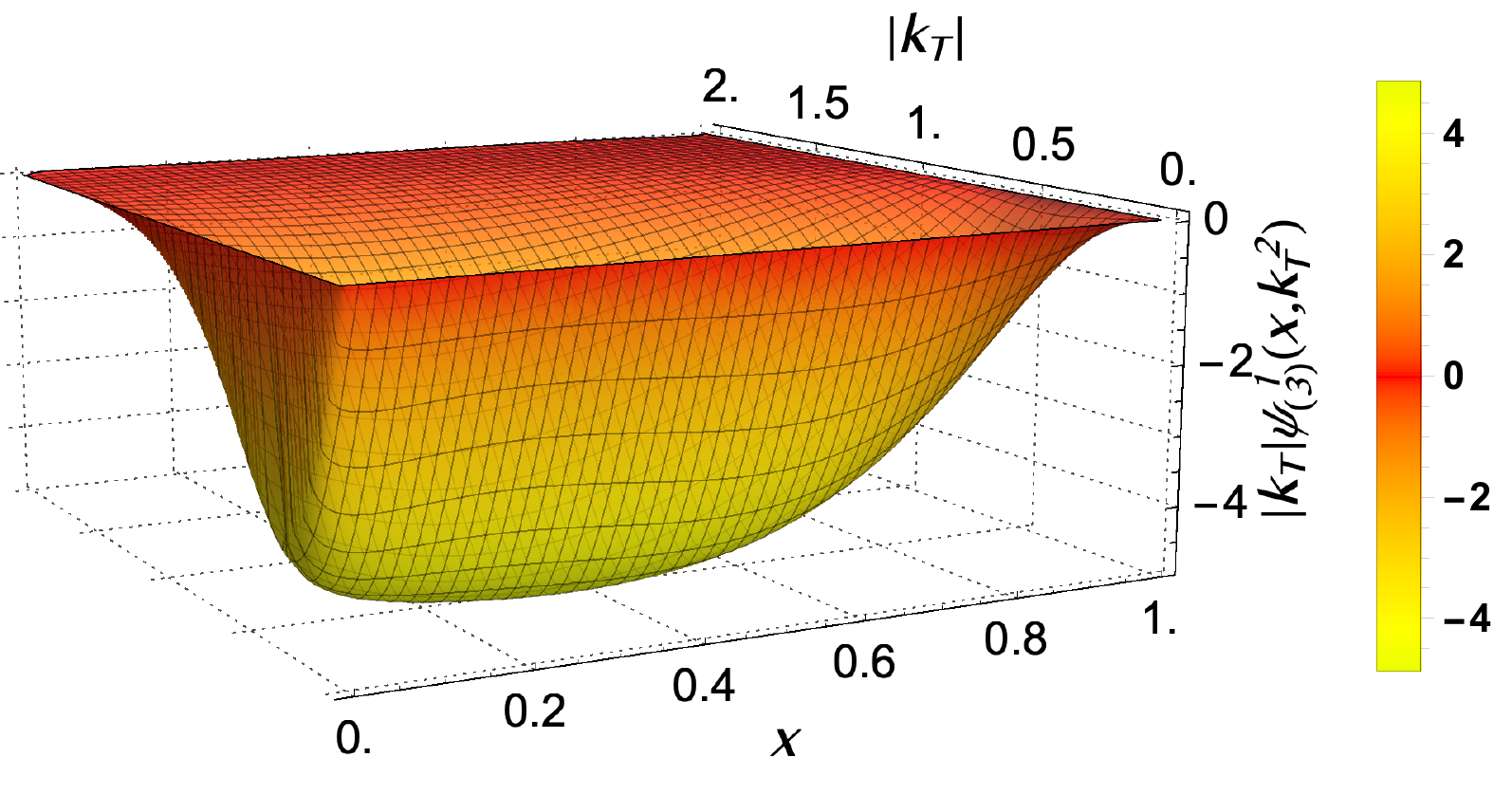}
	\includegraphics[width=3.5in]{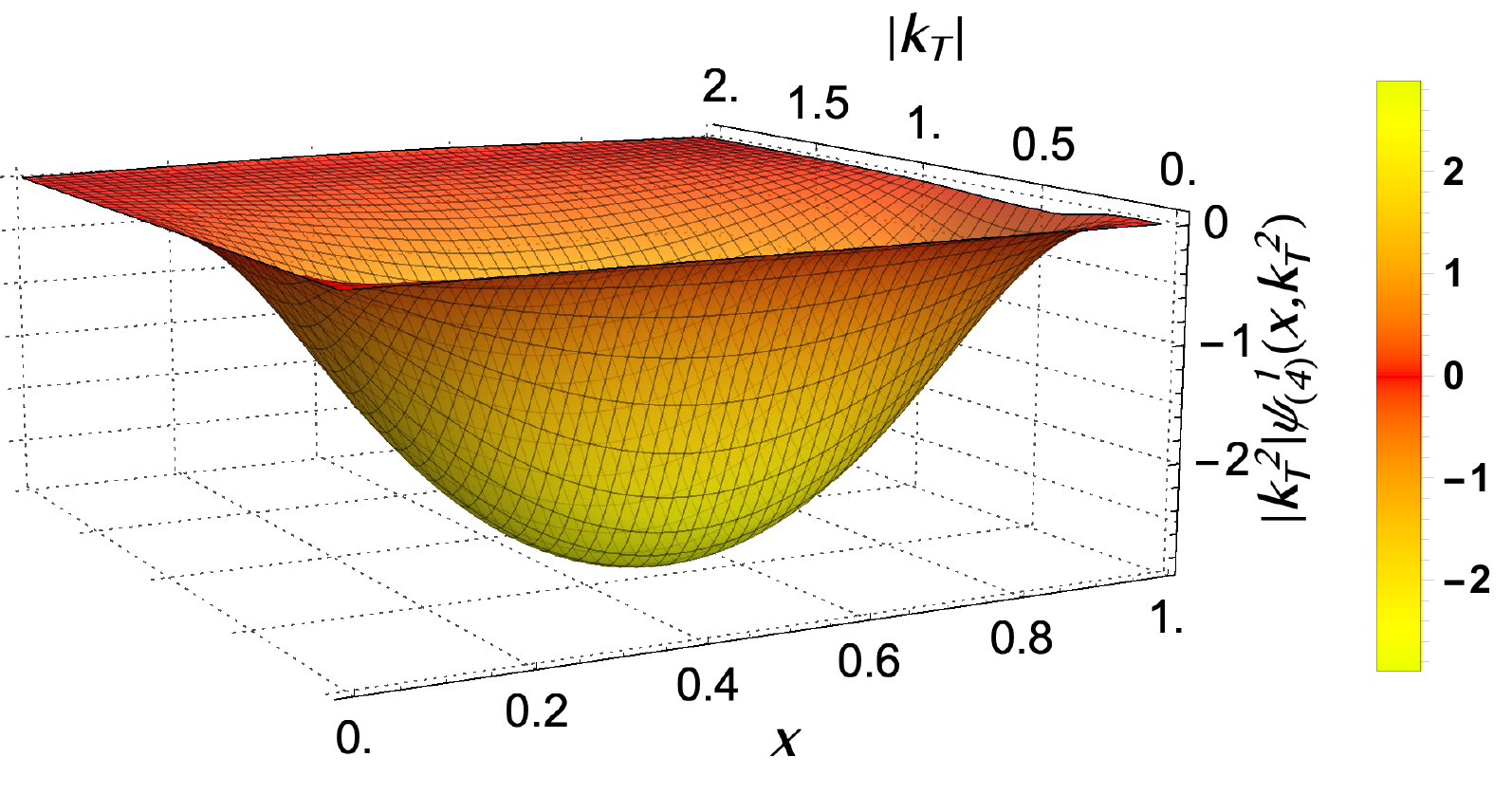}
	\caption{Leading Fock-state LFWFs of longitudinally and transversely polarized $\rho$}
	\label{fig:RhoLFWF}
\end{figure}

\begin{figure}[!t]
	\centering
	\includegraphics[width=3.5in]{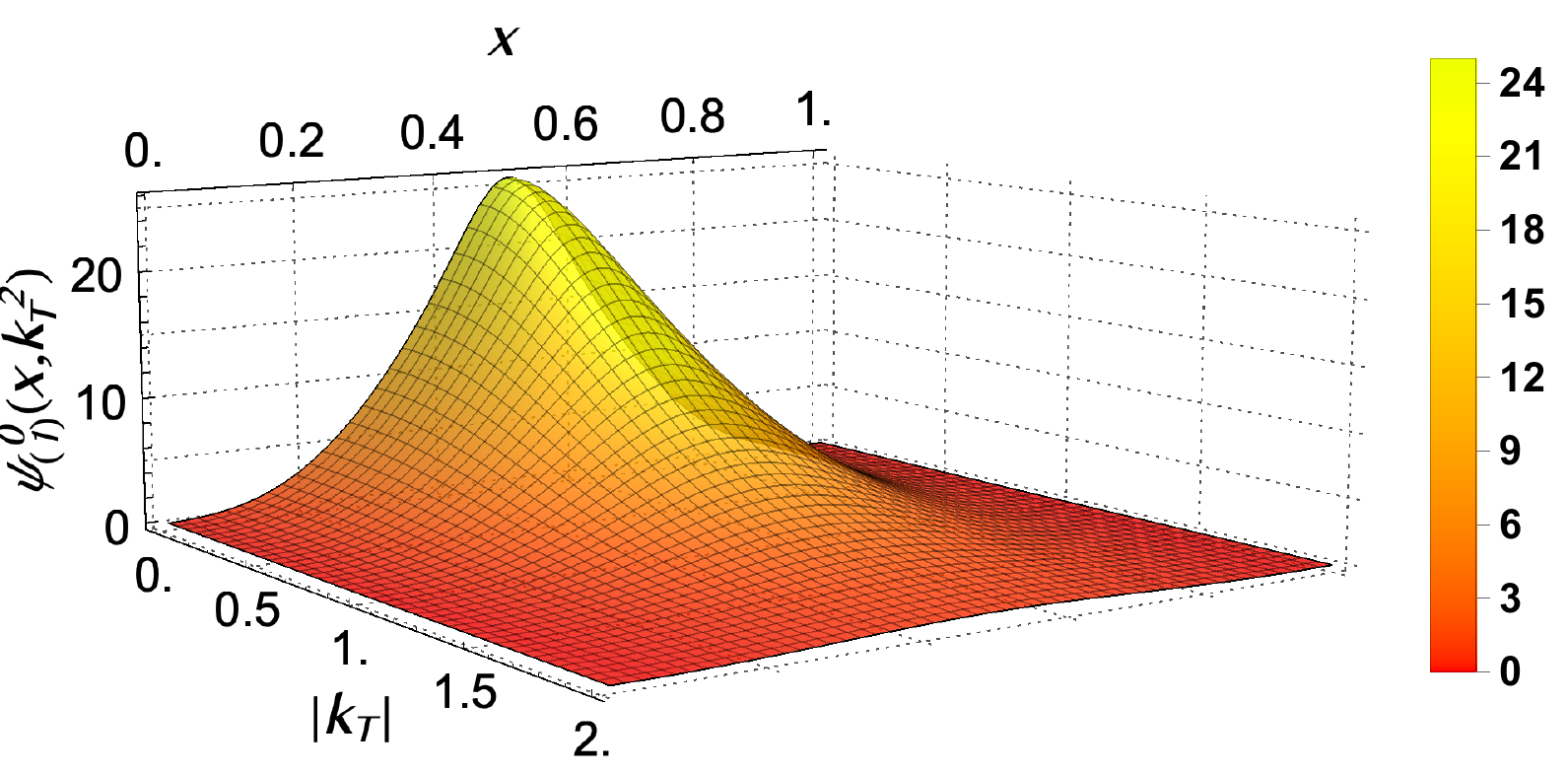}
	\includegraphics[width=3.5in]{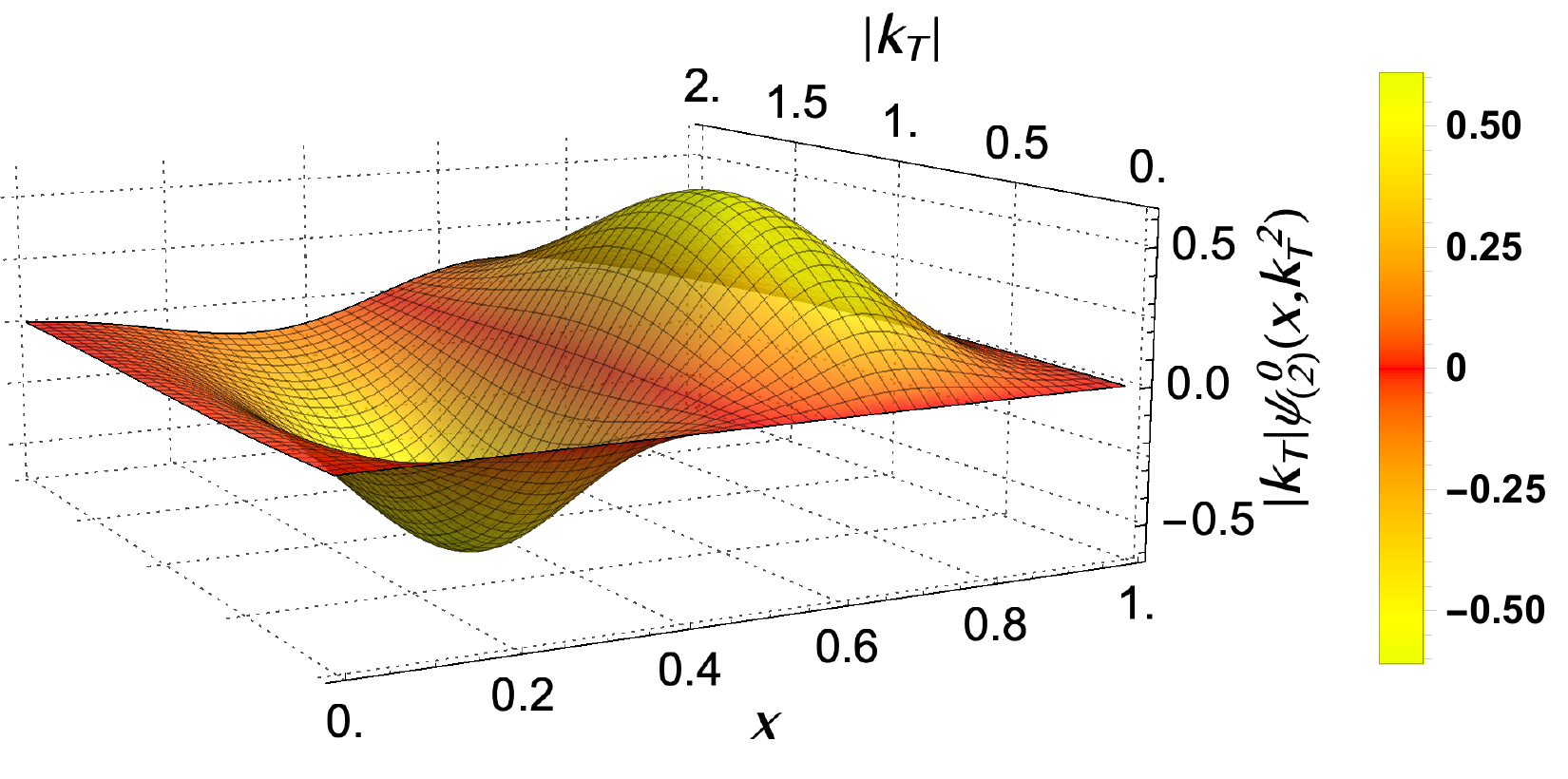}
	\includegraphics[width=3.5in]{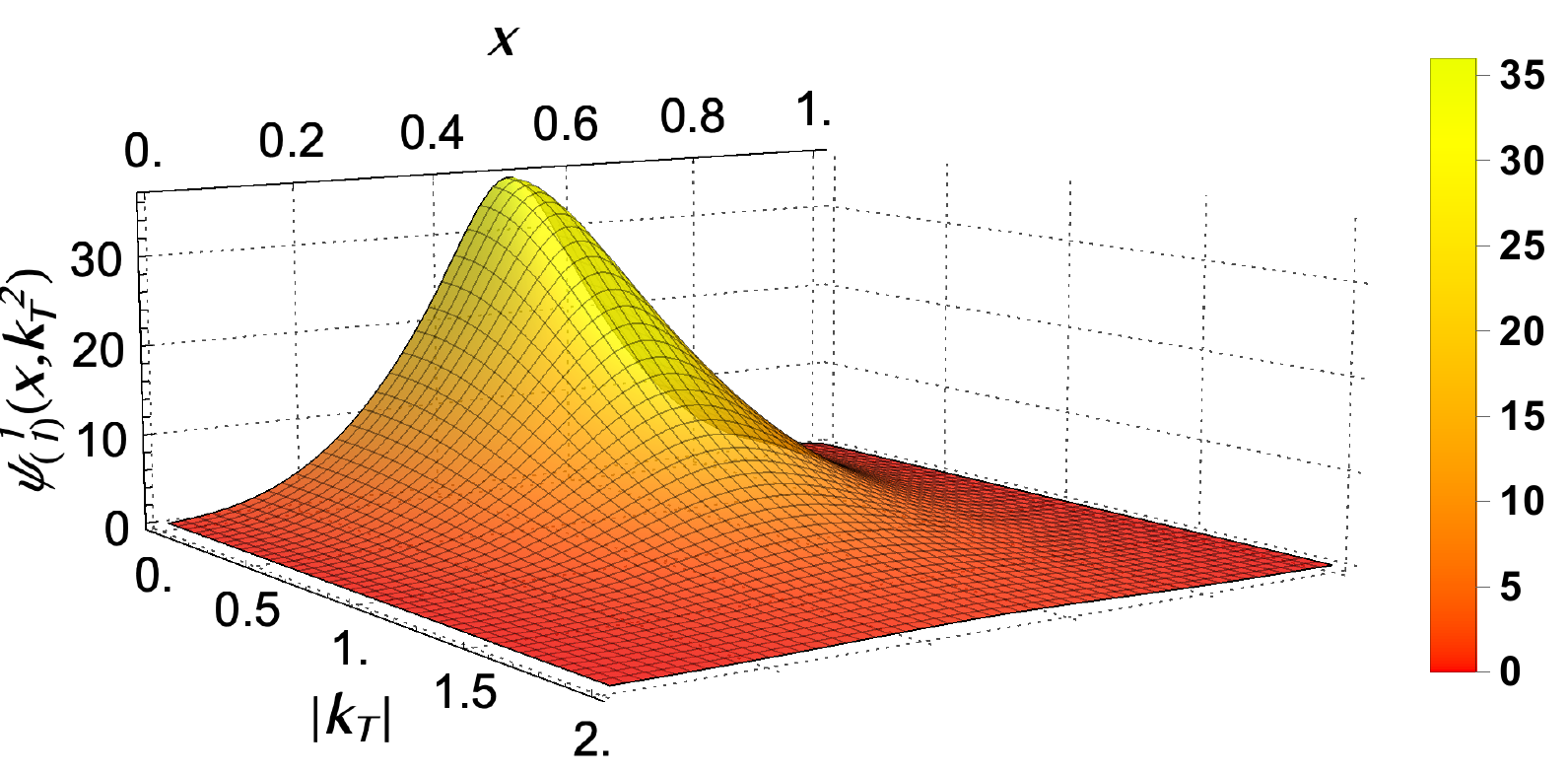}
	\includegraphics[width=3.5in]{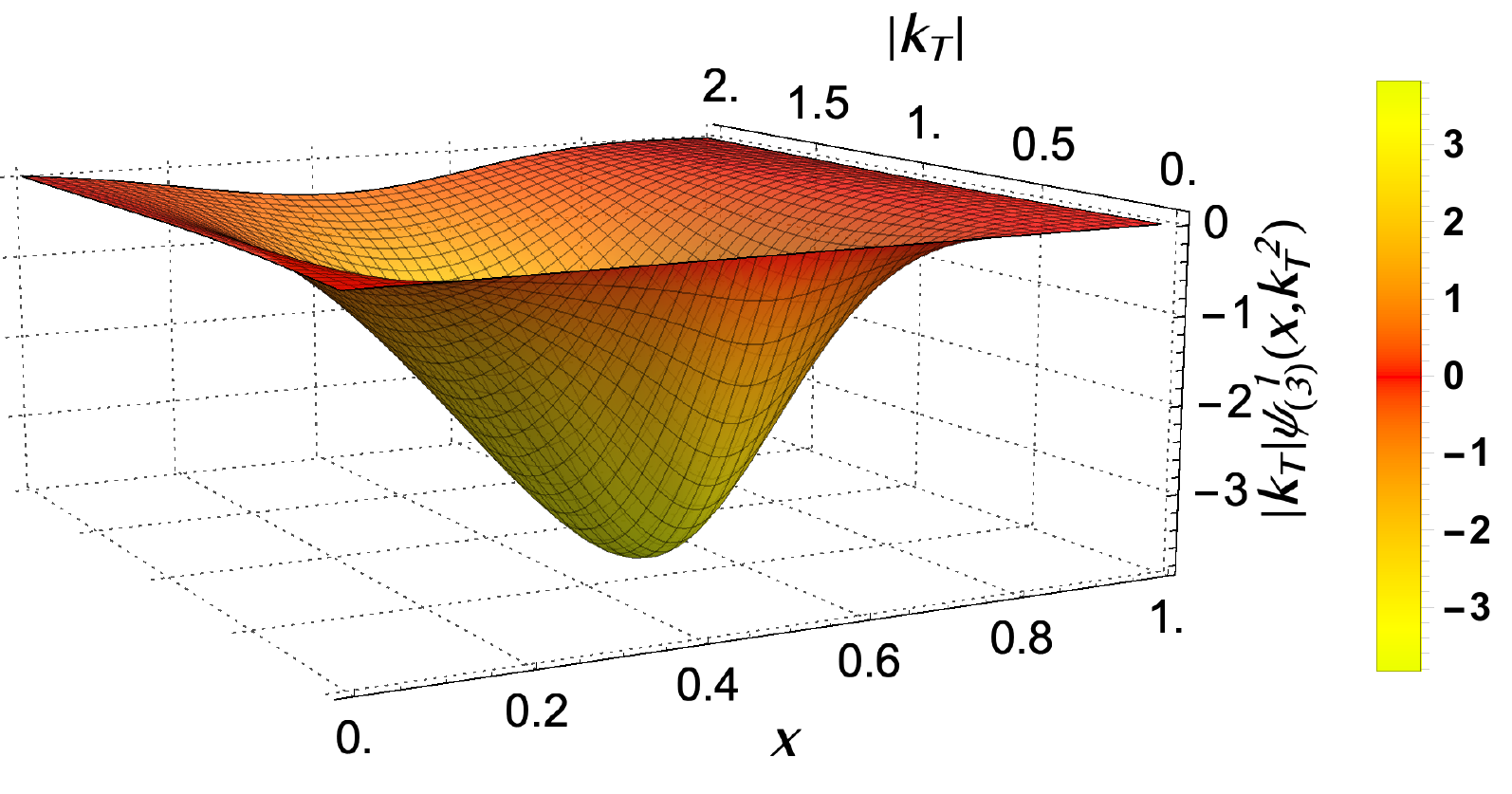}
	\includegraphics[width=3.5in]{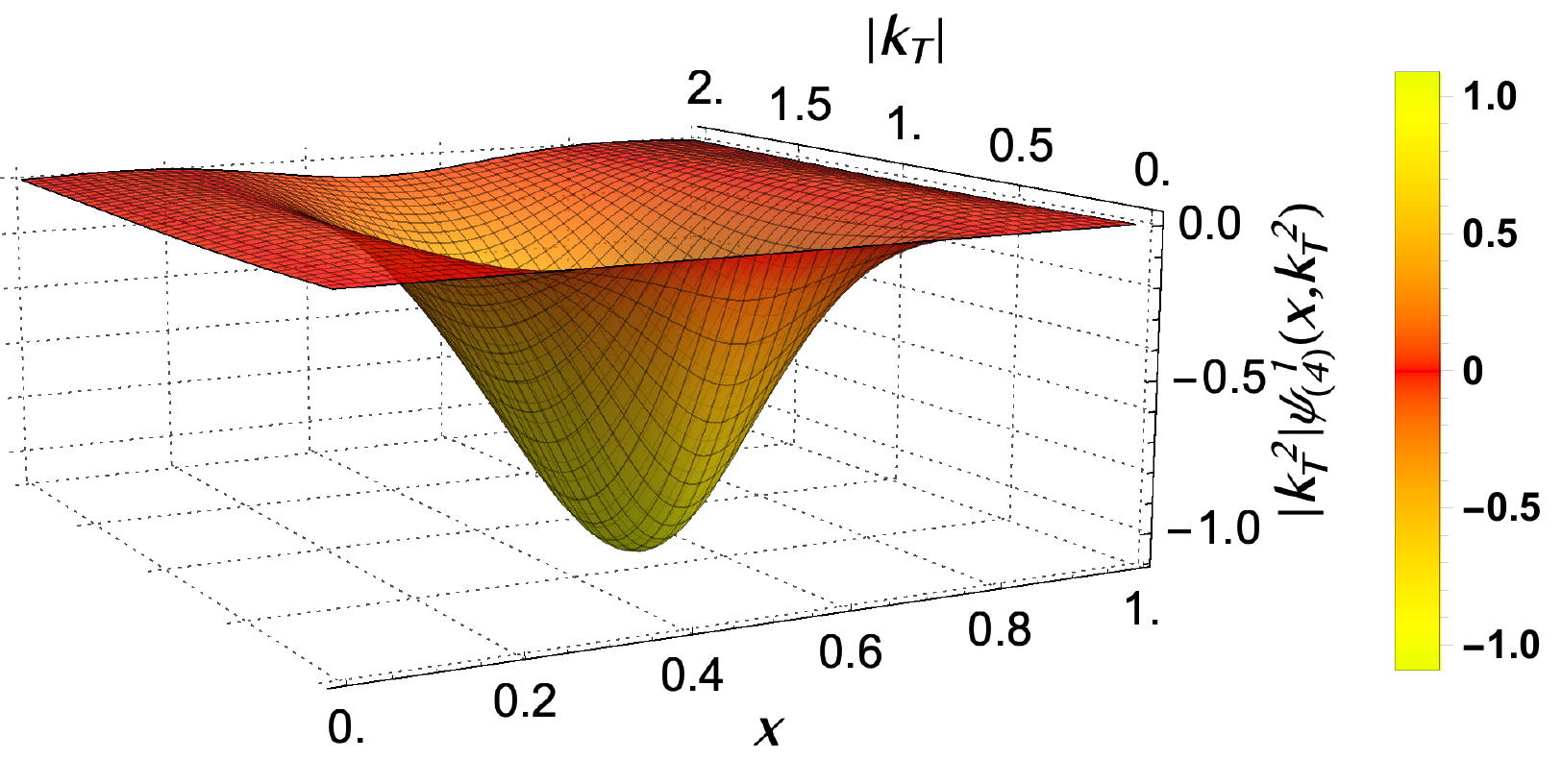}
	\caption{Leading Fock-state LFWFs of longitudinally and transversely polarized $J/\psi$}
	\label{fig:JpsiLFWF}
\end{figure}

We then obtain all five LF-LFWFs of $\rho$ and $J/\psi$, as displayed in Fig.~\ref{fig:RhoLFWF} and Fig.~\ref{fig:JpsiLFWF}. These LF-LFWFs satisfy all the general requirements of Eqs.~(\ref{eq:phiL}-\ref{eq:psi2}). Noticeably, the $\rho$ and $J/\psi$ LFWFs are very different in profile. At small and moderate $\vect{k}_T^2$, the $J/\psi$ LFWFs are distributed closer to $x=1/2$, while the $\rho$ LFWFs are more broadly distributed. This is qualitatively consistent with the phenomenological $\rho$ LFWFs fitted to diffractive $\rho$ production HERA data \cite{Forshaw:2010py} and the AdS/QCD prediction \cite{Forshaw:2012im}. 

A quick comparison of our LF-LFWFs with other theoretical calculations is to look into the twist-2 distribution amplitude (DA) $\phi^V_{\parallel}(x;\mu)$, defined as the $\vect{k}_T-$integrated LFWF
$\phi_{\parallel}^V(x;\mu)=\frac{\sqrt{6}}{f_V}\int^{|\vect{k}_T|=\mu}\frac{d^2 \vect{k}_T}{(2 \pi)^3} \psi_{(1)}^0(x,\vect{k}_T^2).$ For the DA moment $\langle \xi^2 \rangle=\langle (2x-1)^2 \rangle$, we obtain $\langle \xi^2 \rangle^\rho=0.269$ as compared to sum rule results 0.251(24) \cite{Ball:2007zt}, 0.216(21) \cite{Pimikov:2013usa} and 0.241(28) \cite{Fu:2016yzx} at the scale of about $1$ GeV. Note that we determine our scale for $\rho$ to be $\mu \approx 2\omega=1$ GeV, as it is an implicit cutoff within QC-IR model. Meanwhile the lattice QCD gives $\langle \xi^2 \rangle^\rho=0.268(54)$ \cite{Arthur:2010xf} and $0.245(9)$ \cite{Braun:2016wnx} at a higher scale of $2$ GeV. For $J/\psi$, we obtain $\langle \xi^2 \rangle^{J/\psi}=0.093$ as compared to sum rule results $0.083(12)$ \cite{Fu:2018vap} and $0.070(7)$ \cite{Braguta:2007fh} and light-front holography prediction $0.096(20)$ \cite{Li:2017mlw} at the scale of $\mu=m_c$, indicating a significantly narrower DA. 

Beside the broadness, another difference between the $\rho$ and $J/\psi$ LF-LFWFs is their contribution to Fock-states normalization. The meson's LFWFs of all Fock-states should normalize to unity in general, i.e., 
\begin{align}
1&=\sum_{\lambda,\lambda'} N_{\lambda,\lambda'}^\Lambda +N_{\textrm{HF}}, \\
N_{\lambda,\lambda'}^\Lambda&=\int_0^1 dx \int \frac{d \vect{k}_T^2}{2(2 \pi)^3}  |\Phi^\Lambda_{\lambda,\lambda'}(x,\vect{k_T})|^2. \label{eq:N2}
\end{align}
The HF refers to higher Fock-states. Our result is listed in Table.~\ref{tab:N}. The $N_{HF}$ is obtained by subtracting unity with the leading Fock-state contribution. As the DS-BSEs incorporate many higher Fock-states by summing up infinitely many Feynman diagrams, one can see the higher Fock-states contribute considerably to $\rho$ as compared to $J/\psi$. Combining that our $\rho$ LFWFs start with a small current quark mass $m_f=5$ MeV, they hence direct to the parton nature of light quarks inside $\rho$.

\begin{table}[h!]

\begin{center}
\begin{tabular*}
{\hsize}
{
l@{\extracolsep{0ptplus1fil}}
c@{\extracolsep{0ptplus1fil}}
c@{\extracolsep{0ptplus1fil}}
c@{\extracolsep{6ptplus1fil}}
c@{\extracolsep{0ptplus1fil}}
c@{\extracolsep{0ptplus1fil}}
c@{\extracolsep{0ptplus1fil}}
c@{\extracolsep{0ptplus1fil}}
c@{\extracolsep{0ptplus1fil}}
c@{\extracolsep{0ptplus1fil}}
c@{\extracolsep{0ptplus1fil}}}\hline
   & $N_{\uparrow,\downarrow}$ & $N_{\downarrow,\uparrow}$ & $N_{\uparrow,\uparrow}$& $N_{\downarrow,\downarrow}$ & $N_{HF}$   \\\hline
$\rho \ \ \ \ (\Lambda=0)$ & 0.19 & 0.19 & 0.04 & 0.04  & 0.54  \\
$ \ \  \ \ \ \ (\Lambda=1)$ & 0.04 & 0.04 & 0.24 & 0.02  & 0.66  \\
$J/\psi \ (\Lambda=0)$ & 0.44 & 0.44 & 0.01 & 0.01  & 0.10  \\
$\ \  \ \ \ \ (\Lambda=1)$ & 0.03 & 0.03 & 0.78 & $\approx 0.0$  & 0.16 \\\hline
\end{tabular*}
\end{center}
\vspace*{-4ex}
\caption{LFWFs contribution to Fock-states normalization. See Eq.~(\ref{eq:N2}) for definition of $N$. 
\label{tab:N}
}
\end{table}

  \begin{figure}[t!]
	\centering
	\includegraphics[width=3in]{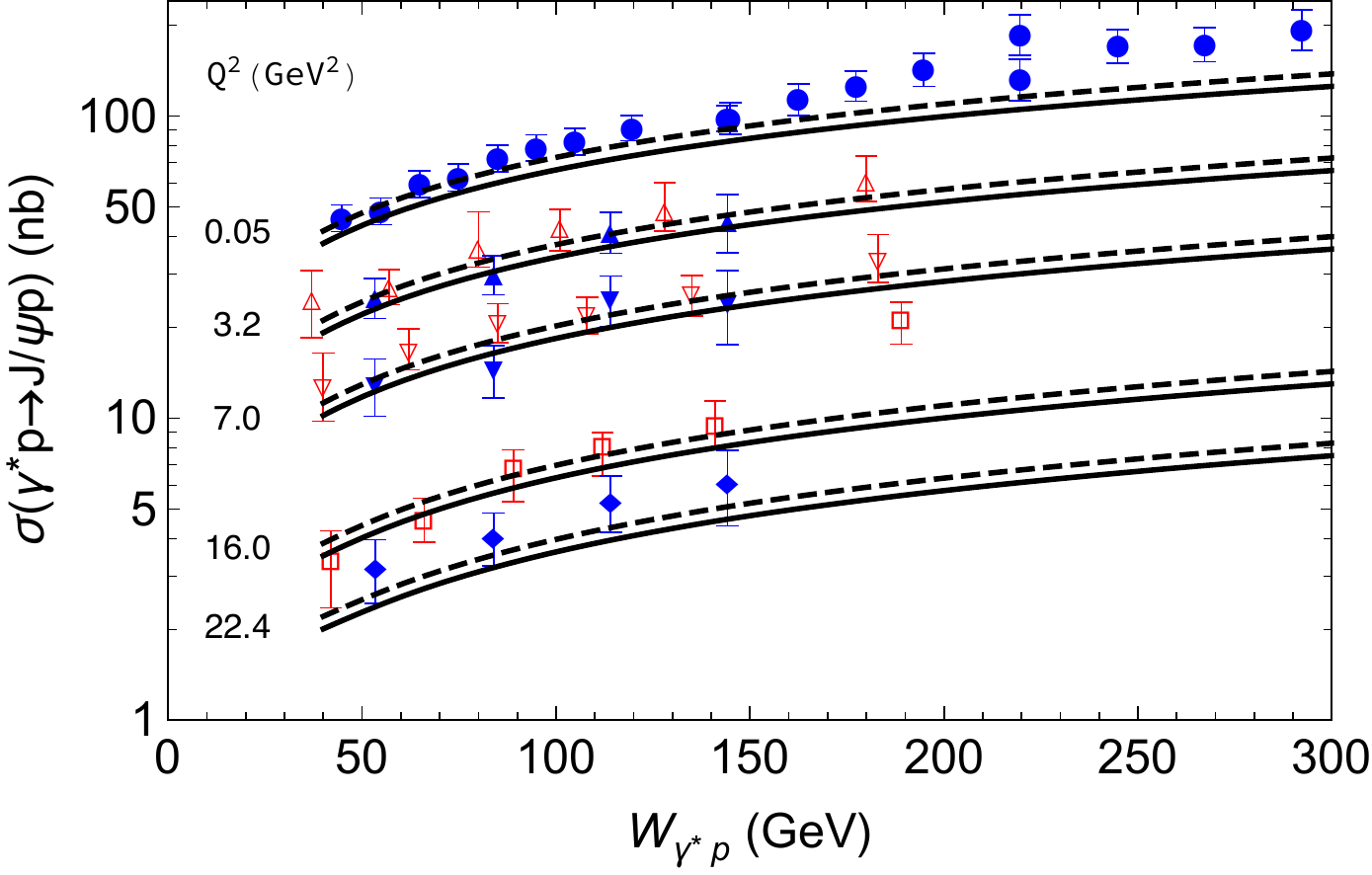}
	\includegraphics[width=3in]{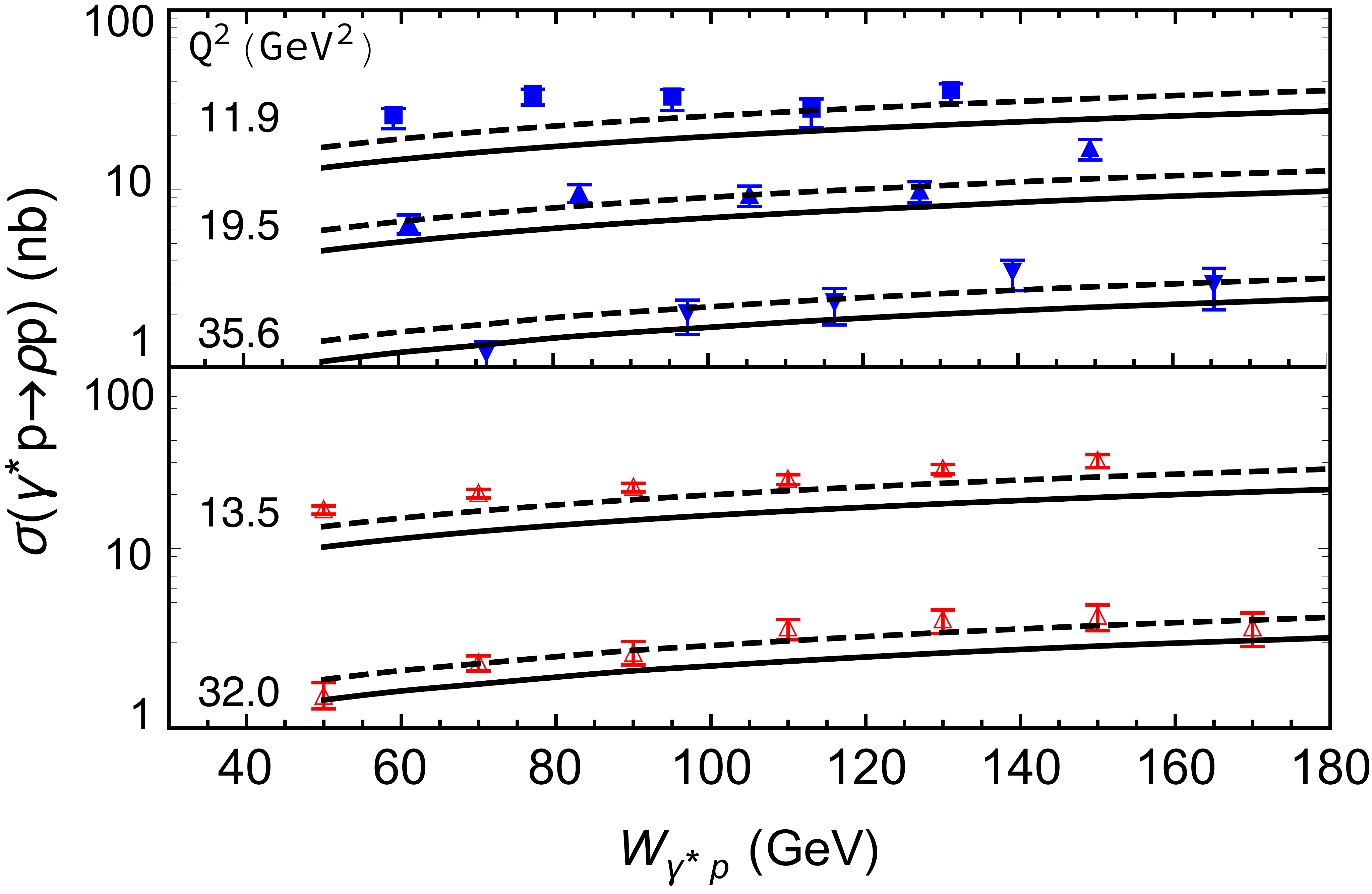}
	\caption{\emph{Upper panel}: Diffractive $J/\psi$ electroproduction cross section obtained using bCGC model and DS-BSEs LF-LFWFs (solid curves). The dashed curves are $1.1$ times the solid curves. 
	The data is taken from H1 \cite{Aktas:2005xu} (filled markers) and ZEUS \cite{Chekanov:2004mw} (empty markers). Note the selected ZEUS data is at $Q^2=3.1$ GeV$^2$ and $6.8$ GeV$^2$. \emph{Lower panel}: Results for $\rho$  (solid curves). The dashed curves are $1.3$ times solid curves. 
	The data is taken from H1 \cite{Aaron:2009xp} (filled markers) and ZEUS \cite{Chekanov:2007zr} (empty markers). Deviation grows when $Q^2$ gets lower than 10 GeV$^2$.}
	\label{fig:TotSig}
\end{figure}

  \begin{figure}[t!]
	\centering
	\includegraphics[width=2.82in]{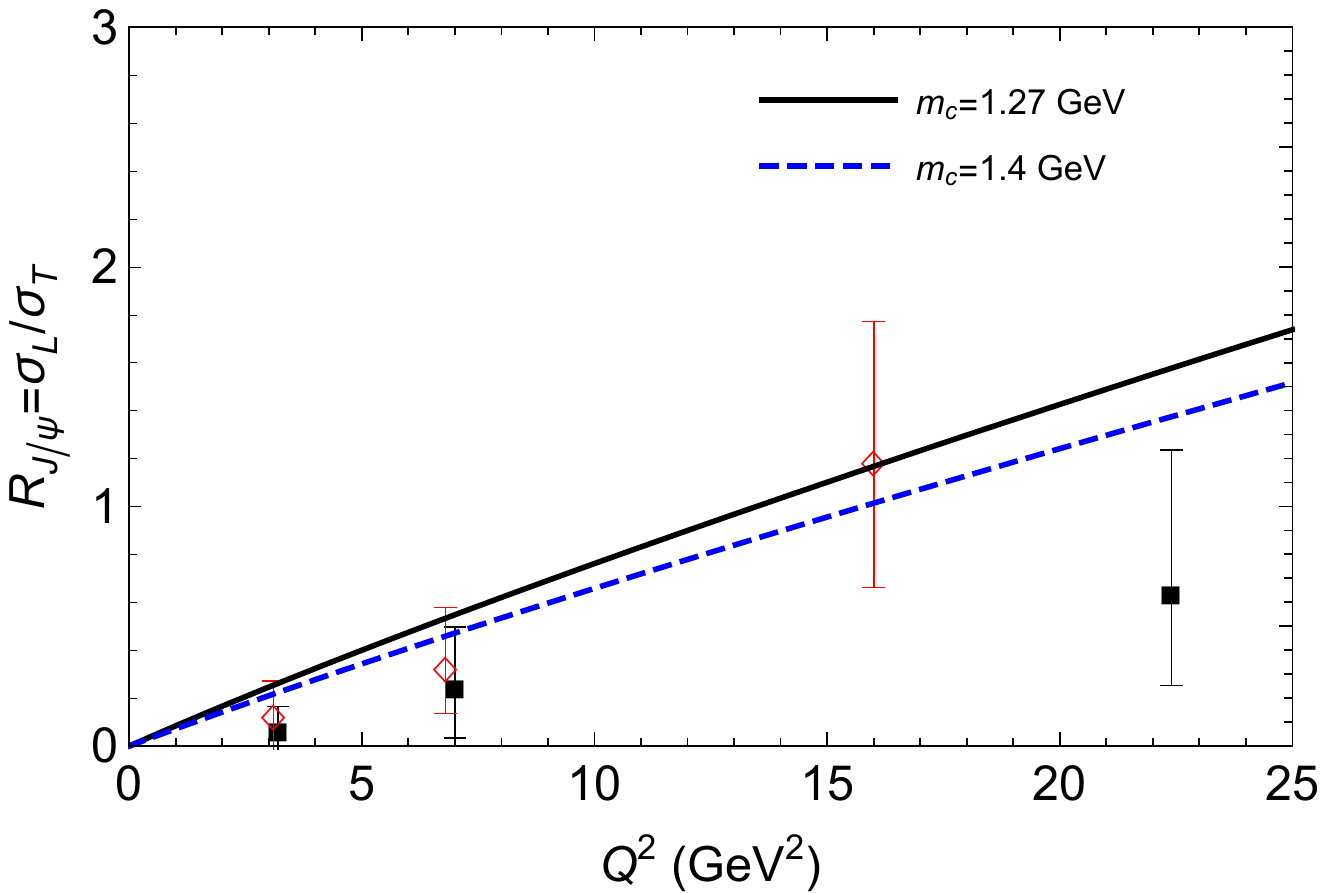}
	\includegraphics[width=2.82in]{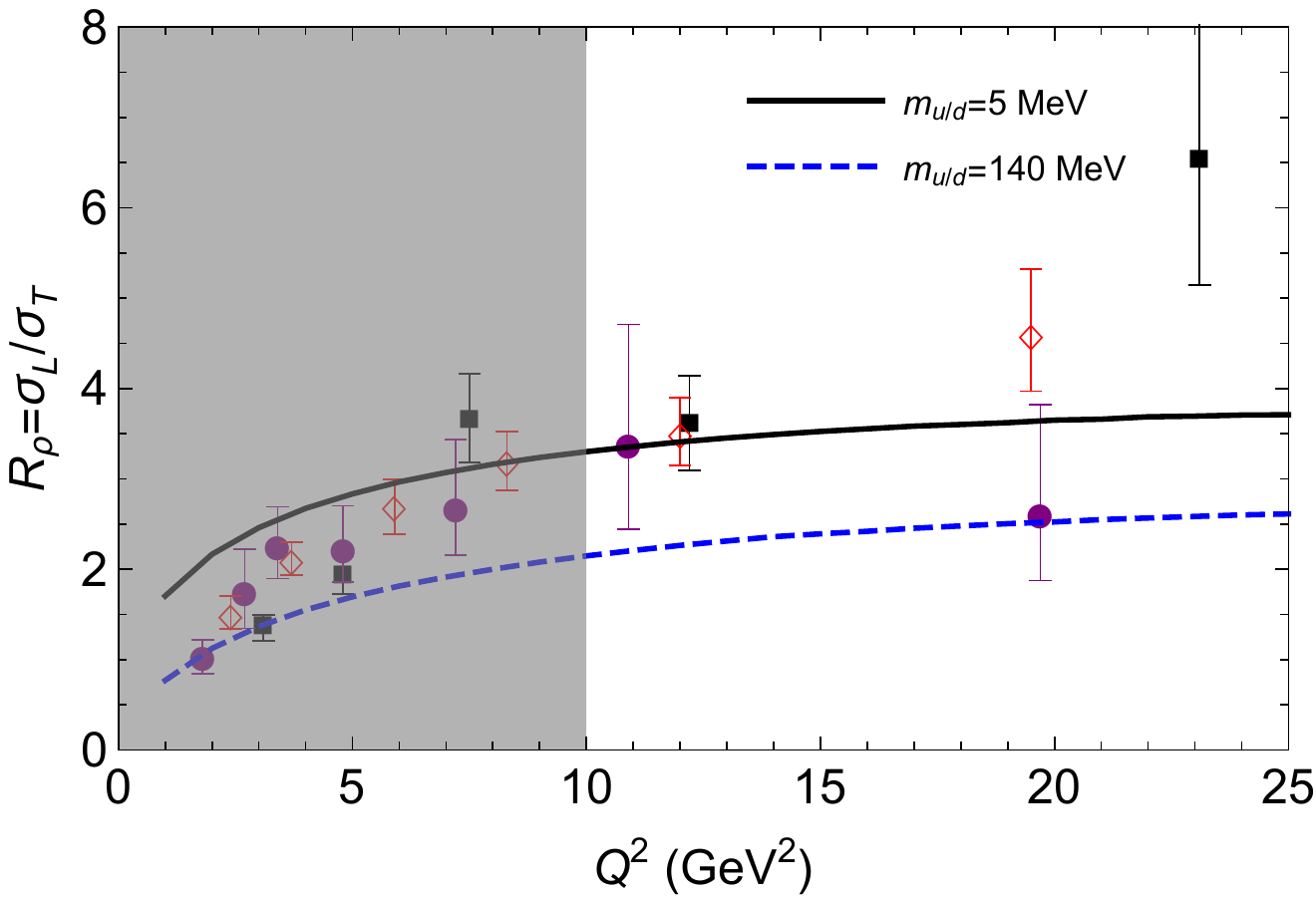}
	\caption{Longitudinal to transverse cross section ratio of $J/\psi$ and $\rho$ production at $W=90$ GeV. \emph{Upper panel}:
$J/\psi$ data taken from H1 \cite{Aktas:2005xu} (filled markers) and ZEUS \cite{Chekanov:2004mw}  (empty markers).  \emph{Lower panel}: $\rho$ data  taken from H1 \cite{Adloff:1999kg, Aaron:2009xp} and ZEUS \cite{Chekanov:2007zr}. The low $Q^2$ region is shaded to indicate where the calculation gets less applicable. Different quark mass parameters are examined in the bCGC dipole model.}
	\label{fig:RatioSig}
\end{figure}

\section{Diffractive $\rho$ and $J/\psi$ electroproduction:} 
Finally we study the diffractive $\rho$ and $J/\psi$ production $\gamma^* p \to Vp$ with the DS-BSEs based LFWFs.  In the dipole picture, the process takes three steps: the virtual photon first splits into a color dipole (quark-anti-quark pair), which then scatters off nucleon via color neutral gluons exchange and finally recombines into the outgoing vector meson, leaving the target nucleon intact \cite{Martin:1999wb,Kowalski:2006hc}.  The scattering amplitude can be factorized into i) the overlap of virtual photon's and vector meson's LF-LFWFs and ii) the amplitude of dipole scattering off a nucleon. Here we follow exactly the conventions and formulas  from section two of \cite{Xie:2018ufg}, with one exception of a minor revision for phase factor proposed in \cite{Hatta:2017cte} (see Eq.~(1) of \cite{Lappi:2020ufv} for the revised form).

Concerning the photon LF-LFWFs, here we employ the leading order QED result  \cite{Dosch:1996ss}
\begin{align}
 \psi_{\lambda\bar{\lambda}, \Lambda= 0}(Q^2)&= -e_fe\sqrt{N_c}\delta_{\lambda,-\bar{\lambda}}2Qx(1-x)\frac{K_0(\epsilon r)}{2\pi},\label{eq:photon1}\\
\psi_{\lambda\bar{\lambda},\Lambda= \pm }(Q^2)& = e_f e\sqrt{2N_c}\left\{ m_f \delta_{\lambda, \pm}\delta_{\bar{\lambda} , \pm}  +  ie^{\pm i\theta_r} [ \mp x\delta_{\lambda, \pm}\delta_{\bar{\lambda} , \mp}
\pm (1-x)\delta_{\lambda, \mp}\delta_{\bar{\lambda} , \pm} ]\partial_r \right \} \frac{K_0(\epsilon r)}{2\pi} ,\label{eq:photon2}
\end{align}
with the photon virtuality $Q^2$, the quark mass $m_f$ and $\epsilon=\sqrt{x(1-x)Q^2+m_f^2}$.  Their form in the $\vect{k}_T$-space can be obtained by Fourier transform with respect to the transverse separation $\vect{r}=(r \textrm{cos}\theta_r, r\textrm{sin}\theta_r)$. They were originally derived within light cone perturbation theory, with loop corrections available in \cite{Beuf:2016wdz,Hanninen:2017ddy}. Here we remark that Eqs.~(\ref{eq:photon1},\ref{eq:photon2}) can also be derived using our method, i.e., calculating Eq.~(\ref{eq:chi2phi}) with the bare quark propagator and quark-photon vertex. Naturally, they can be refined by employing the full quark propagator $S$ and vertex $\Gamma_\mu$, as the solved $S$ and $\Gamma_\mu$ of RL DS-BSEs exhibit considerable dressing effect \cite{Maris:1999bh}. In another word, the photon splitting into light $q\bar{q}$ pair contains not only QED, but also essentially nonperturbative QCD interactions. Such study is ongoing within our effort. Nevertheless, at large $Q^2$ and/or $m_f$ the DCSB effect weakens and the dressed propagator and vertex tend to bare ones. Therefore Eqs.~(\ref{eq:photon1},\ref{eq:photon2})  provide a better approximation in the heavy sector, or in the light sector with relatively high $Q^2$. Meanwhile, Eqs.~(\ref{eq:photon1},\ref{eq:photon2}) inspired some vector meson LFWFs models, such as the Boosted Gaussian model  and Gaus-LC model \cite{Kowalski:2003hm,Kowalski:2006hc,Lappi:2010dd,Rezaeian:2012ji,Xie:2016ino,Xie:2017mil,Xie:2018rog}. Their photon-like parameterization satisfies Eqs.~(\ref{eq:phiL}-\ref{eq:psi2}), but can not fully accommodate the DS-BSEs LFWFs as we checked.

As for the dipole-proton scattering amplitude, there were many successful models \cite{GolecBiernat:1998js,Forshaw:1999uf,Iancu:2003ge,Kowalski:2006hc}. Here we adopt the bCGC model \cite{Iancu:2003ge,Kowalski:2006hc} same as in \cite{Xie:2018ufg}, i.e., with the model parameters originally determined in \cite{Rezaeian:2013tka}. Note that  in analyzing the updated combined HERA small-$x$ DIS data, the bCGC model favors the current light quark mass $m_{u/d}=[10^{-4},10^{-2}]$ GeV \cite{Rezaeian:2013tka}, and hence reveals light quarks' parton nature in small-x diffractive DIS. Here we choose $m_c=1.27$ GeV and $m_{u/d}=5$ MeV, unless otherwise mentioned.

In Fig.~{\ref{fig:TotSig}} we show the $\gamma^* p$ center-of-mass energy ($W$) dependence of the total cross section $\sigma$ for fixed $Q^2$. The upper panel shows our result for $J/\psi$ (solid curves). They generally lie within error bars. As pointed out in \cite{Boer:2011fh,Lappi:2020ufv}, there could be an up to 50\% theoretical uncertainty in the overall normalization of the cross section, which originates from the real to imaginary part of the scattering amplitude ratio correction and in particular the skewedness correction. We therefore multiply all the solid curves by a factor of $1.1$ and get the dashed curves which show better overall agreement. 

The $\rho$ production poses a greater challenge. Since the DS-BSEs LF-LFWFs only contribute less than 50\% to the total normalization, they are significantly smaller in magnitude as compared to phenomenological wave functions that omit higher Fock-states in $\rho$.  Meanwhile as aforementioned, there is larger uncertainty in the virtual photon LF-LFWFs concerning $\gamma^* \rightarrow q\bar{q}$ as compared to $\gamma^* \rightarrow c\bar{c}$ due to nonperturbative effects at low $Q^2$ \cite{Forshaw:2003ki,Berger:2012wx,Goncalves:2020cir}. In practice, we find agreement with HEAR data for $Q^2 \gtrsim 10$ GeV$^2$, as shown in the lower panel of Fig.~\ref{fig:TotSig}. The deviation from data shows up as $Q^2$ gets lower to around $10$ GeV$^2$ and keeps growing. For instance at $Q^2=3.3$ GeV$^2$, the data points are about twice our calculation result.

The longitudinal to transverse cross section ratio doesn't suffer from the absolute normalization uncertainty. We compare HERA data with our calculation in Fig.~\ref{fig:RatioSig}. The quark mass dependence is also examined. Agreement is found in the case of $J/\psi$. We also find a clear preference of small light quark mass $m_{f}=5$ MeV against phenomenological mass $m_{f}=140$ MeV, revealing the parton nature of light (anti)quarks.

\section{Summary and Outlook:} 
 We determine the $\rho$ and $J/\psi$ LF-LFWFs within parton picture by means of the DS-BSEs approach. Employing the color dipole approach and without introducing any new parameters, these LFWFs well reproduce the diffractive $\rho$ and $J/\psi$ electroproduction data at HERA. This work therefore reveals the parton nature of light quarks in the diffractive $\rho$ production. This study can be naturally  extended to the eA collisions at future EIC. Simulations (within dipole approach) in the White Paper \cite{Accardi:2012qut} suggest that i) the diffractive vector meson electroproductions in ep and eA collisions provide good observables for discriminating between saturation and nonsaturation phenomenon and ii) the lighter vector mesons, such as $\rho$ and $\phi$, are more sensitive probes for gluon saturation. Given the large theoretical uncertainties long lying within light vector meson LFWFs, this work therefore paves the way for their diffractive production simulations at EIC (and potentially LHeC \cite{Agostini:2020fmq} and EicC \cite{Anderle:2021wcy}) in the parton basis.

\section{Acknowledgement:} 
We thank Tobias Frederico, Wen-Bao Jia, C\'edric Mezrag, Craig D. Roberts, Peter C. Tandy and Fan Wang for beneficial communications. C.S. also thanks Ian C. Clo\"et for the help in initiating this project. This work is supported by the National Natural Science Foundation of China (under Grant No. 11905104) and the Strategic Priority Research Program of Chinese Academy of Sciences (Grant NO. XDB34030301).

\section{Appendix}
The dressed quark propagator can be generally decomposed as $S(k)=-i \slashed{k}\sigma_V(k^2)+I_4 \sigma_S(k^2)$, as well as the BS amplitude $\Gamma^M_\mu(k,P)=\sum_{i=1}^{8} T^i_\mu(k,P) F^i(k^2,k\cdot P, P^2)$ \cite{Maris:1997hd,Maris:1999nt}. The
 $\sigma_{v/s}$ and $F^i$ are  scalar functions numerically determined  by solving the DS-BSEs. Denoting $A_1=k_\mu-\frac{P_\mu P \cdot k}{P^2}$, $A_2=\gamma_\mu-\frac{P_\mu \slashed{P}}{P^2}$, $B_1=I_4$, $B_2=\slashed{P}$, $B_3=\slashed{k}$ and $B_4=[\slashed{k},\slashed{P}]$, the $T^i_\mu(k,P)$ takes the form
\begin{align}\label{Basis_VC}
T_\mu^1&=i A_1.B_1,     &T_\mu^2&=A_1.B_2 (k \cdot P),  \notag   \\
 T_\mu^3&=A_1.B_3, &T_\mu^4&=-i A_1.B_4,     \notag        \\
 T_\mu^5&=A_2.B_1,      &T_\mu^6 &=-i A_2.B_2,   \notag  \\
 T_\mu^7&=-i[A_2,B_3](k\cdot P),      &T_\mu^8 &=\{A_2, B_4\}\,. 
\end{align}
They are all transverse to vector meson total momentum $P_\mu$ and form a complete Dirac basis for $\Gamma_\mu^M(k,P)$. With such choice, the scalar functions $F^i$ are even in $k\cdot P$ due to negative charge parity of vector mesons.

\begin{table}[hpb]

\begin{center}
\begin{tabular*}
{\hsize}
{
@{\extracolsep{0ptplus1fil}}
c@{\extracolsep{0ptplus1fil}}
c@{\extracolsep{0ptplus1fil}}
c@{\extracolsep{0ptplus1fil}}
c@{\extracolsep{0ptplus1fil}}
c@{\extracolsep{0ptplus1fil}}}\hline
  & $z_1$ & $m_1$  & $z_2$ & $m_2$ \\
charm &    $(0.49, 0.64)$ & $(1.85, 0.55)$& $(0.04, 0.02)$ & $(-2.09,0.85)$  \\
\hline
\end{tabular*}

\begin{tabular*}
{\hsize}
{
l@{\extracolsep{0ptplus1fil}}
c@{\extracolsep{0ptplus1fil}}
c@{\extracolsep{0ptplus1fil}}
c@{\extracolsep{0ptplus1fil}}
c@{\extracolsep{0ptplus1fil}}
c@{\extracolsep{0ptplus1fil}}
c@{\extracolsep{0ptplus1fil}}
c@{\extracolsep{0ptplus1fil}}
c@{\extracolsep{0ptplus1fil}}
c@{\extracolsep{0ptplus1fil}}
c@{\extracolsep{0ptplus1fil}}}\hline
   & $U_1$ & $U_2$ & $U_3$ &$\sigma_1$ & $\sigma_2$ & $\Lambda_1/\Lambda_2$  \\\hline
F$^1$ & 1.83 & -1.22 & 0.08   & -1.86 & -1.9 & 2.2 \\
F$^2$ & -0.079 & 0.082 & 0.0   & -2.83 & -2.63 & 1.8 \\
F$^3$ &0.64 & -0.54 & 0.05   & -2.59 & -2.45 & 1.8 \\
F$^4$ & 0.108 & -0.086 & 0.008   & -2.35 & -2.2 & 2 \\
F$^5$ & 1.43 & -0.49 & 0.06  & -0.006 & 2.31 & 2.4 \\
F$^6$ &-0.277 & 0.281 & 0.0  & -2.59 & -2.5 & 2.2 \\
F$^7$ & 0.435 & -0.406 & 0.01  & -2.79 & -2.62 & 1.8 \\
F$^8$ & 0.131 & -0.079 & 0.01 & -1.45 & -1.27 & 2.2 \\\hline
\end{tabular*}
\end{center}
\vspace*{-4ex}
\caption{Representation parameters. \emph{Upper panel}: Eq.~(\ref{eq:spara}) -- The pair $(x,y)$ represents the complex number $x+ i y$.  \emph{Lower panel}: Eqs~(\ref{eq:fpara}) -- For all eight $F^i$'s we have $n_1=5, n_2=6, n_3=2, \Lambda_1=\Lambda_2, \Lambda_3=1.0$ and $\sigma^3_2=0.0$, with one exception of $n_3=1$ for $F^5$. 
\label{Table:parameters}
}
\end{table}

The fully dressed quark propagator $S(k)$ is then fitted with the sum of pairs of complex conjugate poles \cite{Souchlas:2010boa}
\begin{align}
\label{eq:spara}
S(k)=\sum_{i=1}^{2}\left [ \frac{z_i}{i \sh{k}+m_i}+\frac{z^*_i}{i \sh{k}+m^*_i} \right ],
\end{align}
with parameters in the upper panel of Table.~\ref{Table:parameters}. The scalar functions $F^i(k^2,k\cdot P, P^2)$ of $J/\psi$'s Bethe-Salpeter amplitude are fitted with the Nakanishi-like representation \cite{Nakanishi:1963zz}
\begin{align}
\label{eq:fpara}
F(k;P)&=\sum_{j=1}^3 \left[\int_{-1}^1 d\alpha \rho^{j}(\alpha)\bigg[\frac{U_j \Lambda_j^{2 n_j}}{(k^2+\alpha k\cdot P+\Lambda_j^2)^{n_j}}\bigg] \right ],\\
\rho^j(\alpha)&=\frac{\Gamma(1)}{\Gamma(3/2) \sqrt{\pi}}[C_0^{(1/2)}(\alpha)+\sigma^j_2 C_2^{(1/2)}(\alpha)] 
\label{eq:rho}.
\end{align}
The $C_n^{(1/2)}$ is the Gegenbauer polynomial of order $1/2$. The value of the parameters can be found in the lower panel of Table.~\ref{Table:parameters} and its caption. The method to obtain the point-wisely accurate LFWFs with Eq.~(\ref{eq:chi2phi}) and Eqs.~(\ref{eq:spara}-\ref{eq:rho}) is rather technical. It involves i) analytical computation of traces and Feynman integrals, ii) transform of integration variable and iii) numerical computation of the point-wise behavior of LFWFs. One may refer to \cite{Shi:2018zqd,Shi:2020pqe} for more details.

\bibliography{VM}

\begin{thebibliography}{76}%
\makeatletter
\providecommand \@ifxundefined [1]{%
 \@ifx{#1\undefined}
}%
\providecommand \@ifnum [1]{%
 \ifnum #1\expandafter \@firstoftwo
 \else \expandafter \@secondoftwo
 \fi
}%
\providecommand \@ifx [1]{%
 \ifx #1\expandafter \@firstoftwo
 \else \expandafter \@secondoftwo
 \fi
}%
\providecommand \natexlab [1]{#1}%
\providecommand \enquote  [1]{``#1''}%
\providecommand \bibnamefont  [1]{#1}%
\providecommand \bibfnamefont [1]{#1}%
\providecommand \citenamefont [1]{#1}%
\providecommand \href@noop [0]{\@secondoftwo}%
\providecommand \href [0]{\begingroup \@sanitize@url \@href}%
\providecommand \@href[1]{\@@startlink{#1}\@@href}%
\providecommand \@@href[1]{\endgroup#1\@@endlink}%
\providecommand \@sanitize@url [0]{\catcode `\\12\catcode `\$12\catcode
  `\&12\catcode `\#12\catcode `\^12\catcode `\_12\catcode `\%12\relax}%
\providecommand \@@startlink[1]{}%
\providecommand \@@endlink[0]{}%
\providecommand \url  [0]{\begingroup\@sanitize@url \@url }%
\providecommand \@url [1]{\endgroup\@href {#1}{\urlprefix }}%
\providecommand \urlprefix  [0]{URL }%
\providecommand \Eprint [0]{\href }%
\providecommand \doibase [0]{http://dx.doi.org/}%
\providecommand \selectlanguage [0]{\@gobble}%
\providecommand \bibinfo  [0]{\@secondoftwo}%
\providecommand \bibfield  [0]{\@secondoftwo}%
\providecommand \translation [1]{[#1]}%
\providecommand \BibitemOpen [0]{}%
\providecommand \bibitemStop [0]{}%
\providecommand \bibitemNoStop [0]{.\EOS\space}%
\providecommand \EOS [0]{\spacefactor3000\relax}%
\providecommand \BibitemShut  [1]{\csname bibitem#1\endcsname}%
\let\auto@bib@innerbib\@empty
\bibitem [{\citenamefont {Ryskin}(1993)}]{Ryskin:1992ui}%
  \BibitemOpen
  \bibfield  {author} {\bibinfo {author} {\bibfnamefont {M.}~\bibnamefont
  {Ryskin}},\ }\href {\doibase 10.1007/BF01555742} {\bibfield  {journal}
  {\bibinfo  {journal} {Z. Phys. C}\ }\textbf {\bibinfo {volume} {57}},\
  \bibinfo {pages} {89} (\bibinfo {year} {1993})}\BibitemShut {NoStop}%
\bibitem [{\citenamefont {Armesto}\ and\ \citenamefont
  {Rezaeian}(2014)}]{Armesto:2014sma}%
  \BibitemOpen
  \bibfield  {author} {\bibinfo {author} {\bibfnamefont {N.}~\bibnamefont
  {Armesto}}\ and\ \bibinfo {author} {\bibfnamefont {A.~H.}\ \bibnamefont
  {Rezaeian}},\ }\href {\doibase 10.1103/PhysRevD.90.054003} {\bibfield
  {journal} {\bibinfo  {journal} {Phys. Rev. D}\ }\textbf {\bibinfo {volume}
  {90}},\ \bibinfo {pages} {054003} (\bibinfo {year} {2014})},\ \Eprint
  {http://arxiv.org/abs/1402.4831} {arXiv:1402.4831 [hep-ph]} \BibitemShut
  {NoStop}%
\bibitem [{\citenamefont {Accardi}\ \emph {et~al.}(2016)\citenamefont {Accardi}
  \emph {et~al.}}]{Accardi:2012qut}%
  \BibitemOpen
  \bibfield  {author} {\bibinfo {author} {\bibfnamefont {A.}~\bibnamefont
  {Accardi}} \emph {et~al.},\ }\href {\doibase 10.1140/epja/i2016-16268-9}
  {\bibfield  {journal} {\bibinfo  {journal} {Eur. Phys. J. A}\ }\textbf
  {\bibinfo {volume} {52}},\ \bibinfo {pages} {268} (\bibinfo {year} {2016})},\
  \Eprint {http://arxiv.org/abs/1212.1701} {arXiv:1212.1701 [nucl-ex]}
  \BibitemShut {NoStop}%
\bibitem [{\citenamefont {Brodsky}\ \emph {et~al.}(1994)\citenamefont
  {Brodsky}, \citenamefont {Frankfurt}, \citenamefont {Gunion}, \citenamefont
  {Mueller},\ and\ \citenamefont {Strikman}}]{Brodsky:1994kf}%
  \BibitemOpen
  \bibfield  {author} {\bibinfo {author} {\bibfnamefont {S.~J.}\ \bibnamefont
  {Brodsky}}, \bibinfo {author} {\bibfnamefont {L.}~\bibnamefont {Frankfurt}},
  \bibinfo {author} {\bibfnamefont {J.}~\bibnamefont {Gunion}}, \bibinfo
  {author} {\bibfnamefont {A.~H.}\ \bibnamefont {Mueller}}, \ and\ \bibinfo
  {author} {\bibfnamefont {M.}~\bibnamefont {Strikman}},\ }\href {\doibase
  10.1103/PhysRevD.50.3134} {\bibfield  {journal} {\bibinfo  {journal} {Phys.
  Rev. D}\ }\textbf {\bibinfo {volume} {50}},\ \bibinfo {pages} {3134}
  (\bibinfo {year} {1994})},\ \Eprint {http://arxiv.org/abs/hep-ph/9402283}
  {arXiv:hep-ph/9402283} \BibitemShut {NoStop}%
\bibitem [{\citenamefont {Lappi}\ \emph {et~al.}(2020)\citenamefont {Lappi},
  \citenamefont {M\"antysaari},\ and\ \citenamefont
  {Penttala}}]{Lappi:2020ufv}%
  \BibitemOpen
  \bibfield  {author} {\bibinfo {author} {\bibfnamefont {T.}~\bibnamefont
  {Lappi}}, \bibinfo {author} {\bibfnamefont {H.}~\bibnamefont {M\"antysaari}},
  \ and\ \bibinfo {author} {\bibfnamefont {J.}~\bibnamefont {Penttala}},\
  }\href {\doibase 10.1103/PhysRevD.102.054020} {\bibfield  {journal} {\bibinfo
   {journal} {Phys. Rev. D}\ }\textbf {\bibinfo {volume} {102}},\ \bibinfo
  {pages} {054020} (\bibinfo {year} {2020})},\ \Eprint
  {http://arxiv.org/abs/2006.02830} {arXiv:2006.02830 [hep-ph]} \BibitemShut
  {NoStop}%
\bibitem [{\citenamefont {Brodsky}\ \emph {et~al.}(1998)\citenamefont
  {Brodsky}, \citenamefont {Pauli},\ and\ \citenamefont
  {Pinsky}}]{Brodsky:1997de}%
  \BibitemOpen
  \bibfield  {author} {\bibinfo {author} {\bibfnamefont {S.~J.}\ \bibnamefont
  {Brodsky}}, \bibinfo {author} {\bibfnamefont {H.-C.}\ \bibnamefont {Pauli}},
  \ and\ \bibinfo {author} {\bibfnamefont {S.~S.}\ \bibnamefont {Pinsky}},\
  }\href {\doibase 10.1016/S0370-1573(97)00089-6} {\bibfield  {journal}
  {\bibinfo  {journal} {Phys. Rept.}\ }\textbf {\bibinfo {volume} {301}},\
  \bibinfo {pages} {299} (\bibinfo {year} {1998})},\ \Eprint
  {http://arxiv.org/abs/hep-ph/9705477} {arXiv:hep-ph/9705477} \BibitemShut
  {NoStop}%
\bibitem [{\citenamefont {Kowalski}\ and\ \citenamefont
  {Teaney}(2003)}]{Kowalski:2003hm}%
  \BibitemOpen
  \bibfield  {author} {\bibinfo {author} {\bibfnamefont {H.}~\bibnamefont
  {Kowalski}}\ and\ \bibinfo {author} {\bibfnamefont {D.}~\bibnamefont
  {Teaney}},\ }\href {\doibase 10.1103/PhysRevD.68.114005} {\bibfield
  {journal} {\bibinfo  {journal} {Phys. Rev.}\ }\textbf {\bibinfo {volume}
  {D68}},\ \bibinfo {pages} {114005} (\bibinfo {year} {2003})},\ \Eprint
  {http://arxiv.org/abs/hep-ph/0304189} {arXiv:hep-ph/0304189 [hep-ph]}
  \BibitemShut {NoStop}%
\bibitem [{\citenamefont {Forshaw}\ and\ \citenamefont
  {Sandapen}(2010)}]{Forshaw:2010py}%
  \BibitemOpen
  \bibfield  {author} {\bibinfo {author} {\bibfnamefont {J.}~\bibnamefont
  {Forshaw}}\ and\ \bibinfo {author} {\bibfnamefont {R.}~\bibnamefont
  {Sandapen}},\ }\href {\doibase 10.1007/JHEP11(2010)037} {\bibfield  {journal}
  {\bibinfo  {journal} {JHEP}\ }\textbf {\bibinfo {volume} {11}},\ \bibinfo
  {pages} {037} (\bibinfo {year} {2010})},\ \Eprint
  {http://arxiv.org/abs/1007.1990} {arXiv:1007.1990 [hep-ph]} \BibitemShut
  {NoStop}%
\bibitem [{\citenamefont {Forshaw}\ and\ \citenamefont
  {Sandapen}(2012)}]{Forshaw:2012im}%
  \BibitemOpen
  \bibfield  {author} {\bibinfo {author} {\bibfnamefont {J.}~\bibnamefont
  {Forshaw}}\ and\ \bibinfo {author} {\bibfnamefont {R.}~\bibnamefont
  {Sandapen}},\ }\href {\doibase 10.1103/PhysRevLett.109.081601} {\bibfield
  {journal} {\bibinfo  {journal} {Phys. Rev. Lett.}\ }\textbf {\bibinfo
  {volume} {109}},\ \bibinfo {pages} {081601} (\bibinfo {year} {2012})},\
  \Eprint {http://arxiv.org/abs/1203.6088} {arXiv:1203.6088 [hep-ph]}
  \BibitemShut {NoStop}%
\bibitem [{\citenamefont {Ahmady}\ \emph {et~al.}(2016)\citenamefont {Ahmady},
  \citenamefont {Sandapen},\ and\ \citenamefont {Sharma}}]{Ahmady:2016ujw}%
  \BibitemOpen
  \bibfield  {author} {\bibinfo {author} {\bibfnamefont {M.}~\bibnamefont
  {Ahmady}}, \bibinfo {author} {\bibfnamefont {R.}~\bibnamefont {Sandapen}}, \
  and\ \bibinfo {author} {\bibfnamefont {N.}~\bibnamefont {Sharma}},\ }\href
  {\doibase 10.1103/PhysRevD.94.074018} {\bibfield  {journal} {\bibinfo
  {journal} {Phys. Rev. D}\ }\textbf {\bibinfo {volume} {94}},\ \bibinfo
  {pages} {074018} (\bibinfo {year} {2016})},\ \Eprint
  {http://arxiv.org/abs/1605.07665} {arXiv:1605.07665 [hep-ph]} \BibitemShut
  {NoStop}%
\bibitem [{\citenamefont {'t~Hooft}(1974)}]{tHooft:1974pnl}%
  \BibitemOpen
  \bibfield  {author} {\bibinfo {author} {\bibfnamefont {G.}~\bibnamefont
  {'t~Hooft}},\ }\href {\doibase 10.1016/0550-3213(74)90088-1} {\bibfield
  {journal} {\bibinfo  {journal} {Nucl. Phys. B}\ }\textbf {\bibinfo {volume}
  {75}},\ \bibinfo {pages} {461} (\bibinfo {year} {1974})}\BibitemShut
  {NoStop}%
\bibitem [{\citenamefont {Liu}\ and\ \citenamefont {Soper}(1993)}]{Liu:1992dg}%
  \BibitemOpen
  \bibfield  {author} {\bibinfo {author} {\bibfnamefont {H.~H.}\ \bibnamefont
  {Liu}}\ and\ \bibinfo {author} {\bibfnamefont {D.~E.}\ \bibnamefont
  {Soper}},\ }\href {\doibase 10.1103/PhysRevD.48.1841} {\bibfield  {journal}
  {\bibinfo  {journal} {Phys. Rev. D}\ }\textbf {\bibinfo {volume} {48}},\
  \bibinfo {pages} {1841} (\bibinfo {year} {1993})}\BibitemShut {NoStop}%
\bibitem [{\citenamefont {Burkardt}\ \emph {et~al.}(2002)\citenamefont
  {Burkardt}, \citenamefont {Ji},\ and\ \citenamefont
  {Yuan}}]{Burkardt:2002uc}%
  \BibitemOpen
  \bibfield  {author} {\bibinfo {author} {\bibfnamefont {M.}~\bibnamefont
  {Burkardt}}, \bibinfo {author} {\bibfnamefont {X.-d.}\ \bibnamefont {Ji}}, \
  and\ \bibinfo {author} {\bibfnamefont {F.}~\bibnamefont {Yuan}},\ }\href
  {\doibase 10.1016/S0370-2693(02)02596-0} {\bibfield  {journal} {\bibinfo
  {journal} {Phys. Lett. B}\ }\textbf {\bibinfo {volume} {545}},\ \bibinfo
  {pages} {345} (\bibinfo {year} {2002})},\ \Eprint
  {http://arxiv.org/abs/hep-ph/0205272} {arXiv:hep-ph/0205272} \BibitemShut
  {NoStop}%
\bibitem [{\citenamefont {Roberts}\ and\ \citenamefont
  {Williams}(1994)}]{Roberts:1994dr}%
  \BibitemOpen
  \bibfield  {author} {\bibinfo {author} {\bibfnamefont {C.~D.}\ \bibnamefont
  {Roberts}}\ and\ \bibinfo {author} {\bibfnamefont {A.~G.}\ \bibnamefont
  {Williams}},\ }\href {\doibase 10.1016/0146-6410(94)90049-3} {\bibfield
  {journal} {\bibinfo  {journal} {Prog. Part. Nucl. Phys.}\ }\textbf {\bibinfo
  {volume} {33}},\ \bibinfo {pages} {477} (\bibinfo {year} {1994})},\ \Eprint
  {http://arxiv.org/abs/hep-ph/9403224} {arXiv:hep-ph/9403224 [hep-ph]}
  \BibitemShut {NoStop}%
\bibitem [{\citenamefont {Bashir}\ \emph {et~al.}(2012)\citenamefont {Bashir},
  \citenamefont {Chang}, \citenamefont {Cloet}, \citenamefont {El-Bennich},
  \citenamefont {Liu}, \citenamefont {Roberts},\ and\ \citenamefont
  {Tandy}}]{Bashir:2012fs}%
  \BibitemOpen
  \bibfield  {author} {\bibinfo {author} {\bibfnamefont {A.}~\bibnamefont
  {Bashir}}, \bibinfo {author} {\bibfnamefont {L.}~\bibnamefont {Chang}},
  \bibinfo {author} {\bibfnamefont {I.~C.}\ \bibnamefont {Cloet}}, \bibinfo
  {author} {\bibfnamefont {B.}~\bibnamefont {El-Bennich}}, \bibinfo {author}
  {\bibfnamefont {Y.-X.}\ \bibnamefont {Liu}}, \bibinfo {author} {\bibfnamefont
  {C.~D.}\ \bibnamefont {Roberts}}, \ and\ \bibinfo {author} {\bibfnamefont
  {P.~C.}\ \bibnamefont {Tandy}},\ }\href {\doibase 10.1088/0253-6102/58/1/16}
  {\bibfield  {journal} {\bibinfo  {journal} {Commun. Theor. Phys.}\ }\textbf
  {\bibinfo {volume} {58}},\ \bibinfo {pages} {79} (\bibinfo {year} {2012})},\
  \Eprint {http://arxiv.org/abs/1201.3366} {arXiv:1201.3366 [nucl-th]}
  \BibitemShut {NoStop}%
\bibitem [{\citenamefont {Mezrag}\ \emph {et~al.}(2016)\citenamefont {Mezrag},
  \citenamefont {Moutarde},\ and\ \citenamefont
  {Rodriguez-Quintero}}]{Mezrag:2016hnp}%
  \BibitemOpen
  \bibfield  {author} {\bibinfo {author} {\bibfnamefont {C.}~\bibnamefont
  {Mezrag}}, \bibinfo {author} {\bibfnamefont {H.}~\bibnamefont {Moutarde}}, \
  and\ \bibinfo {author} {\bibfnamefont {J.}~\bibnamefont
  {Rodriguez-Quintero}},\ }\href {\doibase 10.1007/s00601-016-1119-8}
  {\bibfield  {journal} {\bibinfo  {journal} {Few Body Syst.}\ }\textbf
  {\bibinfo {volume} {57}},\ \bibinfo {pages} {729} (\bibinfo {year} {2016})},\
  \Eprint {http://arxiv.org/abs/1602.07722} {arXiv:1602.07722 [nucl-th]}
  \BibitemShut {NoStop}%
\bibitem [{\citenamefont {Shi}\ and\ \citenamefont
  {Clo\"et}(2019)}]{Shi:2018zqd}%
  \BibitemOpen
  \bibfield  {author} {\bibinfo {author} {\bibfnamefont {C.}~\bibnamefont
  {Shi}}\ and\ \bibinfo {author} {\bibfnamefont {I.~C.}\ \bibnamefont
  {Clo\"et}},\ }\href {\doibase 10.1103/PhysRevLett.122.082301} {\bibfield
  {journal} {\bibinfo  {journal} {Phys. Rev. Lett.}\ }\textbf {\bibinfo
  {volume} {122}},\ \bibinfo {pages} {082301} (\bibinfo {year} {2019})},\
  \Eprint {http://arxiv.org/abs/1806.04799} {arXiv:1806.04799 [nucl-th]}
  \BibitemShut {NoStop}%
\bibitem [{\citenamefont {de~Paula}\ \emph {et~al.}(2021)\citenamefont
  {de~Paula}, \citenamefont {Ydrefors}, \citenamefont {Alvarenga~Nogueira},
  \citenamefont {Frederico},\ and\ \citenamefont {Salm\`e}}]{dePaula:2020qna}%
  \BibitemOpen
  \bibfield  {author} {\bibinfo {author} {\bibfnamefont {W.}~\bibnamefont
  {de~Paula}}, \bibinfo {author} {\bibfnamefont {E.}~\bibnamefont {Ydrefors}},
  \bibinfo {author} {\bibfnamefont {J.~H.}\ \bibnamefont {Alvarenga~Nogueira}},
  \bibinfo {author} {\bibfnamefont {T.}~\bibnamefont {Frederico}}, \ and\
  \bibinfo {author} {\bibfnamefont {G.}~\bibnamefont {Salm\`e}},\ }\href
  {\doibase 10.1103/PhysRevD.103.014002} {\bibfield  {journal} {\bibinfo
  {journal} {Phys. Rev. D}\ }\textbf {\bibinfo {volume} {103}},\ \bibinfo
  {pages} {014002} (\bibinfo {year} {2021})},\ \Eprint
  {http://arxiv.org/abs/2012.04973} {arXiv:2012.04973 [hep-ph]} \BibitemShut
  {NoStop}%
\bibitem [{\citenamefont {Ji}\ \emph {et~al.}(2003)\citenamefont {Ji},
  \citenamefont {Ma},\ and\ \citenamefont {Yuan}}]{Ji:2003fw}%
  \BibitemOpen
  \bibfield  {author} {\bibinfo {author} {\bibfnamefont {X.-d.}\ \bibnamefont
  {Ji}}, \bibinfo {author} {\bibfnamefont {J.-P.}\ \bibnamefont {Ma}}, \ and\
  \bibinfo {author} {\bibfnamefont {F.}~\bibnamefont {Yuan}},\ }\href {\doibase
  10.1103/PhysRevLett.90.241601} {\bibfield  {journal} {\bibinfo  {journal}
  {Phys. Rev. Lett.}\ }\textbf {\bibinfo {volume} {90}},\ \bibinfo {pages}
  {241601} (\bibinfo {year} {2003})},\ \Eprint
  {http://arxiv.org/abs/hep-ph/0301141} {arXiv:hep-ph/0301141 [hep-ph]}
  \BibitemShut {NoStop}%
\bibitem [{\citenamefont {Ji}\ \emph {et~al.}(2004)\citenamefont {Ji},
  \citenamefont {Ma},\ and\ \citenamefont {Yuan}}]{Ji:2003yj}%
  \BibitemOpen
  \bibfield  {author} {\bibinfo {author} {\bibfnamefont {X.-d.}\ \bibnamefont
  {Ji}}, \bibinfo {author} {\bibfnamefont {J.-P.}\ \bibnamefont {Ma}}, \ and\
  \bibinfo {author} {\bibfnamefont {F.}~\bibnamefont {Yuan}},\ }\href {\doibase
  10.1140/epjc/s2003-01563-y} {\bibfield  {journal} {\bibinfo  {journal} {Eur.
  Phys. J.}\ }\textbf {\bibinfo {volume} {C33}},\ \bibinfo {pages} {75}
  (\bibinfo {year} {2004})},\ \Eprint {http://arxiv.org/abs/hep-ph/0304107}
  {arXiv:hep-ph/0304107 [hep-ph]} \BibitemShut {NoStop}%
\bibitem [{\citenamefont {Maris}\ and\ \citenamefont
  {Tandy}(1999)}]{Maris:1999nt}%
  \BibitemOpen
  \bibfield  {author} {\bibinfo {author} {\bibfnamefont {P.}~\bibnamefont
  {Maris}}\ and\ \bibinfo {author} {\bibfnamefont {P.~C.}\ \bibnamefont
  {Tandy}},\ }\href {\doibase 10.1103/PhysRevC.60.055214} {\bibfield  {journal}
  {\bibinfo  {journal} {Phys. Rev.}\ }\textbf {\bibinfo {volume} {C60}},\
  \bibinfo {pages} {055214} (\bibinfo {year} {1999})},\ \Eprint
  {http://arxiv.org/abs/nucl-th/9905056} {arXiv:nucl-th/9905056 [nucl-th]}
  \BibitemShut {NoStop}%
\bibitem [{\citenamefont {Bloch}(2002)}]{Bloch:2002eq}%
  \BibitemOpen
  \bibfield  {author} {\bibinfo {author} {\bibfnamefont {J.~C.}\ \bibnamefont
  {Bloch}},\ }\href {\doibase 10.1103/PhysRevD.66.034032} {\bibfield  {journal}
  {\bibinfo  {journal} {Phys. Rev. D}\ }\textbf {\bibinfo {volume} {66}},\
  \bibinfo {pages} {034032} (\bibinfo {year} {2002})},\ \Eprint
  {http://arxiv.org/abs/hep-ph/0202073} {arXiv:hep-ph/0202073} \BibitemShut
  {NoStop}%
\bibitem [{\citenamefont {Qin}\ \emph {et~al.}(2012)\citenamefont {Qin},
  \citenamefont {Chang}, \citenamefont {Liu}, \citenamefont {Roberts},\ and\
  \citenamefont {Wilson}}]{Qin:2011xq}%
  \BibitemOpen
  \bibfield  {author} {\bibinfo {author} {\bibfnamefont {S.-x.}\ \bibnamefont
  {Qin}}, \bibinfo {author} {\bibfnamefont {L.}~\bibnamefont {Chang}}, \bibinfo
  {author} {\bibfnamefont {Y.-x.}\ \bibnamefont {Liu}}, \bibinfo {author}
  {\bibfnamefont {C.~D.}\ \bibnamefont {Roberts}}, \ and\ \bibinfo {author}
  {\bibfnamefont {D.~J.}\ \bibnamefont {Wilson}},\ }\href {\doibase
  10.1103/PhysRevC.85.035202} {\bibfield  {journal} {\bibinfo  {journal} {Phys.
  Rev. C}\ }\textbf {\bibinfo {volume} {85}},\ \bibinfo {pages} {035202}
  (\bibinfo {year} {2012})},\ \Eprint {http://arxiv.org/abs/1109.3459}
  {arXiv:1109.3459 [nucl-th]} \BibitemShut {NoStop}%
\bibitem [{\citenamefont {Maris}\ and\ \citenamefont
  {Roberts}(1997)}]{Maris:1997tm}%
  \BibitemOpen
  \bibfield  {author} {\bibinfo {author} {\bibfnamefont {P.}~\bibnamefont
  {Maris}}\ and\ \bibinfo {author} {\bibfnamefont {C.~D.}\ \bibnamefont
  {Roberts}},\ }\href {\doibase 10.1103/PhysRevC.56.3369} {\bibfield  {journal}
  {\bibinfo  {journal} {Phys. Rev.}\ }\textbf {\bibinfo {volume} {C56}},\
  \bibinfo {pages} {3369} (\bibinfo {year} {1997})},\ \Eprint
  {http://arxiv.org/abs/nucl-th/9708029} {arXiv:nucl-th/9708029 [nucl-th]}
  \BibitemShut {NoStop}%
\bibitem [{\citenamefont {Maris}\ \emph {et~al.}(1998)\citenamefont {Maris},
  \citenamefont {Roberts},\ and\ \citenamefont {Tandy}}]{Maris:1997hd}%
  \BibitemOpen
  \bibfield  {author} {\bibinfo {author} {\bibfnamefont {P.}~\bibnamefont
  {Maris}}, \bibinfo {author} {\bibfnamefont {C.~D.}\ \bibnamefont {Roberts}},
  \ and\ \bibinfo {author} {\bibfnamefont {P.~C.}\ \bibnamefont {Tandy}},\
  }\href {\doibase 10.1016/S0370-2693(97)01535-9} {\bibfield  {journal}
  {\bibinfo  {journal} {Phys. Lett.}\ }\textbf {\bibinfo {volume} {B420}},\
  \bibinfo {pages} {267} (\bibinfo {year} {1998})},\ \Eprint
  {http://arxiv.org/abs/nucl-th/9707003} {arXiv:nucl-th/9707003 [nucl-th]}
  \BibitemShut {NoStop}%
\bibitem [{\citenamefont {Maris}\ and\ \citenamefont
  {Tandy}(2000{\natexlab{a}})}]{Maris:2000sk}%
  \BibitemOpen
  \bibfield  {author} {\bibinfo {author} {\bibfnamefont {P.}~\bibnamefont
  {Maris}}\ and\ \bibinfo {author} {\bibfnamefont {P.~C.}\ \bibnamefont
  {Tandy}},\ }\href {\doibase 10.1103/PhysRevC.62.055204} {\bibfield  {journal}
  {\bibinfo  {journal} {Phys. Rev.}\ }\textbf {\bibinfo {volume} {C62}},\
  \bibinfo {pages} {055204} (\bibinfo {year} {2000}{\natexlab{a}})},\ \Eprint
  {http://arxiv.org/abs/nucl-th/0005015} {arXiv:nucl-th/0005015 [nucl-th]}
  \BibitemShut {NoStop}%
\bibitem [{\citenamefont {Jarecke}\ \emph {et~al.}(2003)\citenamefont
  {Jarecke}, \citenamefont {Maris},\ and\ \citenamefont
  {Tandy}}]{Jarecke:2002xd}%
  \BibitemOpen
  \bibfield  {author} {\bibinfo {author} {\bibfnamefont {D.}~\bibnamefont
  {Jarecke}}, \bibinfo {author} {\bibfnamefont {P.}~\bibnamefont {Maris}}, \
  and\ \bibinfo {author} {\bibfnamefont {P.~C.}\ \bibnamefont {Tandy}},\ }\href
  {\doibase 10.1103/PhysRevC.67.035202} {\bibfield  {journal} {\bibinfo
  {journal} {Phys. Rev. C}\ }\textbf {\bibinfo {volume} {67}},\ \bibinfo
  {pages} {035202} (\bibinfo {year} {2003})},\ \Eprint
  {http://arxiv.org/abs/nucl-th/0208019} {arXiv:nucl-th/0208019} \BibitemShut
  {NoStop}%
\bibitem [{\citenamefont {Bhagwat}\ and\ \citenamefont
  {Maris}(2008)}]{Bhagwat:2006pu}%
  \BibitemOpen
  \bibfield  {author} {\bibinfo {author} {\bibfnamefont {M.}~\bibnamefont
  {Bhagwat}}\ and\ \bibinfo {author} {\bibfnamefont {P.}~\bibnamefont
  {Maris}},\ }\href {\doibase 10.1103/PhysRevC.77.025203} {\bibfield  {journal}
  {\bibinfo  {journal} {Phys. Rev. C}\ }\textbf {\bibinfo {volume} {77}},\
  \bibinfo {pages} {025203} (\bibinfo {year} {2008})},\ \Eprint
  {http://arxiv.org/abs/nucl-th/0612069} {arXiv:nucl-th/0612069} \BibitemShut
  {NoStop}%
\bibitem [{\citenamefont {Xu}\ \emph {et~al.}(2019)\citenamefont {Xu},
  \citenamefont {Binosi}, \citenamefont {Cui}, \citenamefont {Li},
  \citenamefont {Roberts}, \citenamefont {Xu},\ and\ \citenamefont
  {Zong}}]{Xu:2019ilh}%
  \BibitemOpen
  \bibfield  {author} {\bibinfo {author} {\bibfnamefont {Y.-Z.}\ \bibnamefont
  {Xu}}, \bibinfo {author} {\bibfnamefont {D.}~\bibnamefont {Binosi}}, \bibinfo
  {author} {\bibfnamefont {Z.-F.}\ \bibnamefont {Cui}}, \bibinfo {author}
  {\bibfnamefont {B.-L.}\ \bibnamefont {Li}}, \bibinfo {author} {\bibfnamefont
  {C.~D.}\ \bibnamefont {Roberts}}, \bibinfo {author} {\bibfnamefont {S.-S.}\
  \bibnamefont {Xu}}, \ and\ \bibinfo {author} {\bibfnamefont {H.~S.}\
  \bibnamefont {Zong}},\ }\href {\doibase 10.1103/PhysRevD.100.114038}
  {\bibfield  {journal} {\bibinfo  {journal} {Phys. Rev. D}\ }\textbf {\bibinfo
  {volume} {100}},\ \bibinfo {pages} {114038} (\bibinfo {year} {2019})},\
  \Eprint {http://arxiv.org/abs/1911.05199} {arXiv:1911.05199 [nucl-th]}
  \BibitemShut {NoStop}%
\bibitem [{\citenamefont {Eichmann}\ \emph {et~al.}(2010)\citenamefont
  {Eichmann}, \citenamefont {Alkofer}, \citenamefont {Krassnigg},\ and\
  \citenamefont {Nicmorus}}]{Eichmann:2009qa}%
  \BibitemOpen
  \bibfield  {author} {\bibinfo {author} {\bibfnamefont {G.}~\bibnamefont
  {Eichmann}}, \bibinfo {author} {\bibfnamefont {R.}~\bibnamefont {Alkofer}},
  \bibinfo {author} {\bibfnamefont {A.}~\bibnamefont {Krassnigg}}, \ and\
  \bibinfo {author} {\bibfnamefont {D.}~\bibnamefont {Nicmorus}},\ }\href
  {\doibase 10.1103/PhysRevLett.104.201601} {\bibfield  {journal} {\bibinfo
  {journal} {Phys. Rev. Lett.}\ }\textbf {\bibinfo {volume} {104}},\ \bibinfo
  {pages} {201601} (\bibinfo {year} {2010})},\ \Eprint
  {http://arxiv.org/abs/0912.2246} {arXiv:0912.2246 [hep-ph]} \BibitemShut
  {NoStop}%
\bibitem [{\citenamefont {Eichmann}(2011)}]{Eichmann:2011vu}%
  \BibitemOpen
  \bibfield  {author} {\bibinfo {author} {\bibfnamefont {G.}~\bibnamefont
  {Eichmann}},\ }\href {\doibase 10.1103/PhysRevD.84.014014} {\bibfield
  {journal} {\bibinfo  {journal} {Phys. Rev. D}\ }\textbf {\bibinfo {volume}
  {84}},\ \bibinfo {pages} {014014} (\bibinfo {year} {2011})},\ \Eprint
  {http://arxiv.org/abs/1104.4505} {arXiv:1104.4505 [hep-ph]} \BibitemShut
  {NoStop}%
\bibitem [{\citenamefont {Fischer}\ and\ \citenamefont
  {Williams}(2009)}]{Fischer:2009jm}%
  \BibitemOpen
  \bibfield  {author} {\bibinfo {author} {\bibfnamefont {C.~S.}\ \bibnamefont
  {Fischer}}\ and\ \bibinfo {author} {\bibfnamefont {R.}~\bibnamefont
  {Williams}},\ }\href {\doibase 10.1103/PhysRevLett.103.122001} {\bibfield
  {journal} {\bibinfo  {journal} {Phys. Rev. Lett.}\ }\textbf {\bibinfo
  {volume} {103}},\ \bibinfo {pages} {122001} (\bibinfo {year} {2009})},\
  \Eprint {http://arxiv.org/abs/0905.2291} {arXiv:0905.2291 [hep-ph]}
  \BibitemShut {NoStop}%
\bibitem [{\citenamefont {Chang}\ and\ \citenamefont
  {Roberts}(2009)}]{Chang:2009zb}%
  \BibitemOpen
  \bibfield  {author} {\bibinfo {author} {\bibfnamefont {L.}~\bibnamefont
  {Chang}}\ and\ \bibinfo {author} {\bibfnamefont {C.~D.}\ \bibnamefont
  {Roberts}},\ }\href {\doibase 10.1103/PhysRevLett.103.081601} {\bibfield
  {journal} {\bibinfo  {journal} {Phys. Rev. Lett.}\ }\textbf {\bibinfo
  {volume} {103}},\ \bibinfo {pages} {081601} (\bibinfo {year} {2009})},\
  \Eprint {http://arxiv.org/abs/0903.5461} {arXiv:0903.5461 [nucl-th]}
  \BibitemShut {NoStop}%
\bibitem [{\citenamefont {Shi}\ \emph {et~al.}(2014)\citenamefont {Shi},
  \citenamefont {Chang}, \citenamefont {Roberts}, \citenamefont {Schmidt},
  \citenamefont {Tandy},\ and\ \citenamefont {Zong}}]{Shi:2014uwa}%
  \BibitemOpen
  \bibfield  {author} {\bibinfo {author} {\bibfnamefont {C.}~\bibnamefont
  {Shi}}, \bibinfo {author} {\bibfnamefont {L.}~\bibnamefont {Chang}}, \bibinfo
  {author} {\bibfnamefont {C.~D.}\ \bibnamefont {Roberts}}, \bibinfo {author}
  {\bibfnamefont {S.~M.}\ \bibnamefont {Schmidt}}, \bibinfo {author}
  {\bibfnamefont {P.~C.}\ \bibnamefont {Tandy}}, \ and\ \bibinfo {author}
  {\bibfnamefont {H.-S.}\ \bibnamefont {Zong}},\ }\href {\doibase
  10.1016/j.physletb.2014.07.057} {\bibfield  {journal} {\bibinfo  {journal}
  {Phys. Lett.}\ }\textbf {\bibinfo {volume} {B738}},\ \bibinfo {pages} {512}
  (\bibinfo {year} {2014})},\ \Eprint {http://arxiv.org/abs/1406.3353}
  {arXiv:1406.3353 [nucl-th]} \BibitemShut {NoStop}%
\bibitem [{\citenamefont {Zyla}\ \emph {et~al.}(2020)\citenamefont {Zyla} \emph
  {et~al.}}]{Zyla:2020zbs}%
  \BibitemOpen
  \bibfield  {author} {\bibinfo {author} {\bibfnamefont {P.}~\bibnamefont
  {Zyla}} \emph {et~al.} (\bibinfo {collaboration} {Particle Data Group}),\
  }\href {\doibase 10.1093/ptep/ptaa104} {\bibfield  {journal} {\bibinfo
  {journal} {PTEP}\ }\textbf {\bibinfo {volume} {2020}},\ \bibinfo {pages}
  {083C01} (\bibinfo {year} {2020})}\BibitemShut {NoStop}%
\bibitem [{\citenamefont {Shi}\ \emph {et~al.}(2015)\citenamefont {Shi},
  \citenamefont {Chen}, \citenamefont {Chang}, \citenamefont {Roberts},
  \citenamefont {Schmidt},\ and\ \citenamefont {Zong}}]{Shi:2015esa}%
  \BibitemOpen
  \bibfield  {author} {\bibinfo {author} {\bibfnamefont {C.}~\bibnamefont
  {Shi}}, \bibinfo {author} {\bibfnamefont {C.}~\bibnamefont {Chen}}, \bibinfo
  {author} {\bibfnamefont {L.}~\bibnamefont {Chang}}, \bibinfo {author}
  {\bibfnamefont {C.~D.}\ \bibnamefont {Roberts}}, \bibinfo {author}
  {\bibfnamefont {S.~M.}\ \bibnamefont {Schmidt}}, \ and\ \bibinfo {author}
  {\bibfnamefont {H.-S.}\ \bibnamefont {Zong}},\ }\href {\doibase
  10.1103/PhysRevD.92.014035} {\bibfield  {journal} {\bibinfo  {journal} {Phys.
  Rev. D}\ }\textbf {\bibinfo {volume} {92}},\ \bibinfo {pages} {014035}
  (\bibinfo {year} {2015})},\ \Eprint {http://arxiv.org/abs/1504.00689}
  {arXiv:1504.00689 [nucl-th]} \BibitemShut {NoStop}%
\bibitem [{\citenamefont {Chen}\ \emph {et~al.}(2020)\citenamefont {Chen},
  \citenamefont {Chang},\ and\ \citenamefont {Liu}}]{Chen:2020ecu}%
  \BibitemOpen
  \bibfield  {author} {\bibinfo {author} {\bibfnamefont {M.}~\bibnamefont
  {Chen}}, \bibinfo {author} {\bibfnamefont {L.}~\bibnamefont {Chang}}, \ and\
  \bibinfo {author} {\bibfnamefont {Y.-x.}\ \bibnamefont {Liu}},\ }\href
  {\doibase 10.1103/PhysRevD.101.056002} {\bibfield  {journal} {\bibinfo
  {journal} {Phys. Rev. D}\ }\textbf {\bibinfo {volume} {101}},\ \bibinfo
  {pages} {056002} (\bibinfo {year} {2020})},\ \Eprint
  {http://arxiv.org/abs/2001.00161} {arXiv:2001.00161 [hep-ph]} \BibitemShut
  {NoStop}%
\bibitem [{\citenamefont {Shi}\ \emph {et~al.}(2020)\citenamefont {Shi},
  \citenamefont {Bednar}, \citenamefont {Clo\"et},\ and\ \citenamefont
  {Freese}}]{Shi:2020pqe}%
  \BibitemOpen
  \bibfield  {author} {\bibinfo {author} {\bibfnamefont {C.}~\bibnamefont
  {Shi}}, \bibinfo {author} {\bibfnamefont {K.}~\bibnamefont {Bednar}},
  \bibinfo {author} {\bibfnamefont {I.~C.}\ \bibnamefont {Clo\"et}}, \ and\
  \bibinfo {author} {\bibfnamefont {A.}~\bibnamefont {Freese}},\ }\href
  {\doibase 10.1103/PhysRevD.101.074014} {\bibfield  {journal} {\bibinfo
  {journal} {Phys. Rev. D}\ }\textbf {\bibinfo {volume} {101}},\ \bibinfo
  {pages} {074014} (\bibinfo {year} {2020})},\ \Eprint
  {http://arxiv.org/abs/2003.03037} {arXiv:2003.03037 [hep-ph]} \BibitemShut
  {NoStop}%
\bibitem [{\citenamefont {Nakanishi}(1963)}]{Nakanishi:1963zz}%
  \BibitemOpen
  \bibfield  {author} {\bibinfo {author} {\bibfnamefont {N.}~\bibnamefont
  {Nakanishi}},\ }\href {\doibase 10.1103/PhysRev.130.1230} {\bibfield
  {journal} {\bibinfo  {journal} {Phys. Rev.}\ }\textbf {\bibinfo {volume}
  {130}},\ \bibinfo {pages} {1230} (\bibinfo {year} {1963})}\BibitemShut
  {NoStop}%
\bibitem [{\citenamefont {Souchlas}(2010)}]{Souchlas:2010boa}%
  \BibitemOpen
  \bibfield  {author} {\bibinfo {author} {\bibfnamefont {N.}~\bibnamefont
  {Souchlas}},\ }\href {\doibase 10.1088/0954-3899/37/11/115001} {\bibfield
  {journal} {\bibinfo  {journal} {J. Phys. G}\ }\textbf {\bibinfo {volume}
  {37}},\ \bibinfo {pages} {115001} (\bibinfo {year} {2010})}\BibitemShut
  {NoStop}%
\bibitem [{\citenamefont {Ball}\ \emph {et~al.}(2007)\citenamefont {Ball},
  \citenamefont {Braun},\ and\ \citenamefont {Lenz}}]{Ball:2007zt}%
  \BibitemOpen
  \bibfield  {author} {\bibinfo {author} {\bibfnamefont {P.}~\bibnamefont
  {Ball}}, \bibinfo {author} {\bibfnamefont {V.}~\bibnamefont {Braun}}, \ and\
  \bibinfo {author} {\bibfnamefont {A.}~\bibnamefont {Lenz}},\ }\href {\doibase
  10.1088/1126-6708/2007/08/090} {\bibfield  {journal} {\bibinfo  {journal}
  {JHEP}\ }\textbf {\bibinfo {volume} {08}},\ \bibinfo {pages} {090} (\bibinfo
  {year} {2007})},\ \Eprint {http://arxiv.org/abs/0707.1201} {arXiv:0707.1201
  [hep-ph]} \BibitemShut {NoStop}%
\bibitem [{\citenamefont {Pimikov}\ \emph {et~al.}(2014)\citenamefont
  {Pimikov}, \citenamefont {Mikhailov},\ and\ \citenamefont
  {Stefanis}}]{Pimikov:2013usa}%
  \BibitemOpen
  \bibfield  {author} {\bibinfo {author} {\bibfnamefont {A.}~\bibnamefont
  {Pimikov}}, \bibinfo {author} {\bibfnamefont {S.}~\bibnamefont {Mikhailov}},
  \ and\ \bibinfo {author} {\bibfnamefont {N.}~\bibnamefont {Stefanis}},\
  }\href {\doibase 10.1007/s00601-014-0815-5} {\bibfield  {journal} {\bibinfo
  {journal} {Few Body Syst.}\ }\textbf {\bibinfo {volume} {55}},\ \bibinfo
  {pages} {401} (\bibinfo {year} {2014})},\ \Eprint
  {http://arxiv.org/abs/1312.2776} {arXiv:1312.2776 [hep-ph]} \BibitemShut
  {NoStop}%
\bibitem [{\citenamefont {Fu}\ \emph {et~al.}(2016)\citenamefont {Fu},
  \citenamefont {Wu}, \citenamefont {Cheng},\ and\ \citenamefont
  {Zhong}}]{Fu:2016yzx}%
  \BibitemOpen
  \bibfield  {author} {\bibinfo {author} {\bibfnamefont {H.-B.}\ \bibnamefont
  {Fu}}, \bibinfo {author} {\bibfnamefont {X.-G.}\ \bibnamefont {Wu}}, \bibinfo
  {author} {\bibfnamefont {W.}~\bibnamefont {Cheng}}, \ and\ \bibinfo {author}
  {\bibfnamefont {T.}~\bibnamefont {Zhong}},\ }\href {\doibase
  10.1103/PhysRevD.94.074004} {\bibfield  {journal} {\bibinfo  {journal} {Phys.
  Rev. D}\ }\textbf {\bibinfo {volume} {94}},\ \bibinfo {pages} {074004}
  (\bibinfo {year} {2016})},\ \Eprint {http://arxiv.org/abs/1607.04937}
  {arXiv:1607.04937 [hep-ph]} \BibitemShut {NoStop}%
\bibitem [{\citenamefont {Arthur}\ \emph {et~al.}(2011)\citenamefont {Arthur},
  \citenamefont {Boyle}, \citenamefont {Brommel}, \citenamefont {Donnellan},
  \citenamefont {Flynn}, \citenamefont {Juttner}, \citenamefont {Rae},\ and\
  \citenamefont {Sachrajda}}]{Arthur:2010xf}%
  \BibitemOpen
  \bibfield  {author} {\bibinfo {author} {\bibfnamefont {R.}~\bibnamefont
  {Arthur}}, \bibinfo {author} {\bibfnamefont {P.}~\bibnamefont {Boyle}},
  \bibinfo {author} {\bibfnamefont {D.}~\bibnamefont {Brommel}}, \bibinfo
  {author} {\bibfnamefont {M.}~\bibnamefont {Donnellan}}, \bibinfo {author}
  {\bibfnamefont {J.}~\bibnamefont {Flynn}}, \bibinfo {author} {\bibfnamefont
  {A.}~\bibnamefont {Juttner}}, \bibinfo {author} {\bibfnamefont
  {T.}~\bibnamefont {Rae}}, \ and\ \bibinfo {author} {\bibfnamefont
  {C.}~\bibnamefont {Sachrajda}},\ }\href {\doibase 10.1103/PhysRevD.83.074505}
  {\bibfield  {journal} {\bibinfo  {journal} {Phys. Rev. D}\ }\textbf {\bibinfo
  {volume} {83}},\ \bibinfo {pages} {074505} (\bibinfo {year} {2011})},\
  \Eprint {http://arxiv.org/abs/1011.5906} {arXiv:1011.5906 [hep-lat]}
  \BibitemShut {NoStop}%
\bibitem [{\citenamefont {Braun}\ \emph {et~al.}(2017)\citenamefont {Braun}
  \emph {et~al.}}]{Braun:2016wnx}%
  \BibitemOpen
  \bibfield  {author} {\bibinfo {author} {\bibfnamefont {V.~M.}\ \bibnamefont
  {Braun}} \emph {et~al.},\ }\href {\doibase 10.1007/JHEP04(2017)082}
  {\bibfield  {journal} {\bibinfo  {journal} {JHEP}\ }\textbf {\bibinfo
  {volume} {04}},\ \bibinfo {pages} {082} (\bibinfo {year} {2017})},\ \Eprint
  {http://arxiv.org/abs/1612.02955} {arXiv:1612.02955 [hep-lat]} \BibitemShut
  {NoStop}%
\bibitem [{\citenamefont {Fu}\ \emph {et~al.}(2018)\citenamefont {Fu},
  \citenamefont {Zeng}, \citenamefont {Cheng}, \citenamefont {Wu},\ and\
  \citenamefont {Zhong}}]{Fu:2018vap}%
  \BibitemOpen
  \bibfield  {author} {\bibinfo {author} {\bibfnamefont {H.-B.}\ \bibnamefont
  {Fu}}, \bibinfo {author} {\bibfnamefont {L.}~\bibnamefont {Zeng}}, \bibinfo
  {author} {\bibfnamefont {W.}~\bibnamefont {Cheng}}, \bibinfo {author}
  {\bibfnamefont {X.-G.}\ \bibnamefont {Wu}}, \ and\ \bibinfo {author}
  {\bibfnamefont {T.}~\bibnamefont {Zhong}},\ }\href {\doibase
  10.1103/PhysRevD.97.074025} {\bibfield  {journal} {\bibinfo  {journal} {Phys.
  Rev. D}\ }\textbf {\bibinfo {volume} {97}},\ \bibinfo {pages} {074025}
  (\bibinfo {year} {2018})},\ \Eprint {http://arxiv.org/abs/1801.06832}
  {arXiv:1801.06832 [hep-ph]} \BibitemShut {NoStop}%
\bibitem [{\citenamefont {Braguta}(2007)}]{Braguta:2007fh}%
  \BibitemOpen
  \bibfield  {author} {\bibinfo {author} {\bibfnamefont {V.}~\bibnamefont
  {Braguta}},\ }\href {\doibase 10.1103/PhysRevD.75.094016} {\bibfield
  {journal} {\bibinfo  {journal} {Phys. Rev. D}\ }\textbf {\bibinfo {volume}
  {75}},\ \bibinfo {pages} {094016} (\bibinfo {year} {2007})},\ \Eprint
  {http://arxiv.org/abs/hep-ph/0701234} {arXiv:hep-ph/0701234} \BibitemShut
  {NoStop}%
\bibitem [{\citenamefont {Li}\ \emph {et~al.}(2017)\citenamefont {Li},
  \citenamefont {Maris},\ and\ \citenamefont {Vary}}]{Li:2017mlw}%
  \BibitemOpen
  \bibfield  {author} {\bibinfo {author} {\bibfnamefont {Y.}~\bibnamefont
  {Li}}, \bibinfo {author} {\bibfnamefont {P.}~\bibnamefont {Maris}}, \ and\
  \bibinfo {author} {\bibfnamefont {J.~P.}\ \bibnamefont {Vary}},\ }\href
  {\doibase 10.1103/PhysRevD.96.016022} {\bibfield  {journal} {\bibinfo
  {journal} {Phys. Rev.}\ }\textbf {\bibinfo {volume} {D96}},\ \bibinfo {pages}
  {016022} (\bibinfo {year} {2017})},\ \Eprint
  {http://arxiv.org/abs/1704.06968} {arXiv:1704.06968 [hep-ph]} \BibitemShut
  {NoStop}%
\bibitem [{\citenamefont {Aktas}\ \emph {et~al.}(2006)\citenamefont {Aktas}
  \emph {et~al.}}]{Aktas:2005xu}%
  \BibitemOpen
  \bibfield  {author} {\bibinfo {author} {\bibfnamefont {A.}~\bibnamefont
  {Aktas}} \emph {et~al.} (\bibinfo {collaboration} {H1}),\ }\href {\doibase
  10.1140/epjc/s2006-02519-5} {\bibfield  {journal} {\bibinfo  {journal} {Eur.
  Phys. J. C}\ }\textbf {\bibinfo {volume} {46}},\ \bibinfo {pages} {585}
  (\bibinfo {year} {2006})},\ \Eprint {http://arxiv.org/abs/hep-ex/0510016}
  {arXiv:hep-ex/0510016} \BibitemShut {NoStop}%
\bibitem [{\citenamefont {Chekanov}\ \emph {et~al.}(2004)\citenamefont
  {Chekanov} \emph {et~al.}}]{Chekanov:2004mw}%
  \BibitemOpen
  \bibfield  {author} {\bibinfo {author} {\bibfnamefont {S.}~\bibnamefont
  {Chekanov}} \emph {et~al.} (\bibinfo {collaboration} {ZEUS}),\ }\href
  {\doibase 10.1016/j.nuclphysb.2004.06.034} {\bibfield  {journal} {\bibinfo
  {journal} {Nucl. Phys. B}\ }\textbf {\bibinfo {volume} {695}},\ \bibinfo
  {pages} {3} (\bibinfo {year} {2004})},\ \Eprint
  {http://arxiv.org/abs/hep-ex/0404008} {arXiv:hep-ex/0404008} \BibitemShut
  {NoStop}%
\bibitem [{\citenamefont {Aaron}\ \emph {et~al.}(2010)\citenamefont {Aaron}
  \emph {et~al.}}]{Aaron:2009xp}%
  \BibitemOpen
  \bibfield  {author} {\bibinfo {author} {\bibfnamefont {F.}~\bibnamefont
  {Aaron}} \emph {et~al.} (\bibinfo {collaboration} {H1}),\ }\href {\doibase
  10.1007/JHEP05(2010)032} {\bibfield  {journal} {\bibinfo  {journal} {JHEP}\
  }\textbf {\bibinfo {volume} {05}},\ \bibinfo {pages} {032} (\bibinfo {year}
  {2010})},\ \Eprint {http://arxiv.org/abs/0910.5831} {arXiv:0910.5831
  [hep-ex]} \BibitemShut {NoStop}%
\bibitem [{\citenamefont {Chekanov}\ \emph {et~al.}(2007)\citenamefont
  {Chekanov} \emph {et~al.}}]{Chekanov:2007zr}%
  \BibitemOpen
  \bibfield  {author} {\bibinfo {author} {\bibfnamefont {S.}~\bibnamefont
  {Chekanov}} \emph {et~al.} (\bibinfo {collaboration} {ZEUS}),\ }\href
  {\doibase 10.1186/1754-0410-1-6} {\bibfield  {journal} {\bibinfo  {journal}
  {PMC Phys. A}\ }\textbf {\bibinfo {volume} {1}},\ \bibinfo {pages} {6}
  (\bibinfo {year} {2007})},\ \Eprint {http://arxiv.org/abs/0708.1478}
  {arXiv:0708.1478 [hep-ex]} \BibitemShut {NoStop}%
\bibitem [{\citenamefont {Adloff}\ \emph {et~al.}(2000)\citenamefont {Adloff}
  \emph {et~al.}}]{Adloff:1999kg}%
  \BibitemOpen
  \bibfield  {author} {\bibinfo {author} {\bibfnamefont {C.}~\bibnamefont
  {Adloff}} \emph {et~al.} (\bibinfo {collaboration} {H1}),\ }\href {\doibase
  10.1007/s100520050703} {\bibfield  {journal} {\bibinfo  {journal} {Eur. Phys.
  J. C}\ }\textbf {\bibinfo {volume} {13}},\ \bibinfo {pages} {371} (\bibinfo
  {year} {2000})},\ \Eprint {http://arxiv.org/abs/hep-ex/9902019}
  {arXiv:hep-ex/9902019} \BibitemShut {NoStop}%
\bibitem [{\citenamefont {Martin}\ \emph {et~al.}(2000)\citenamefont {Martin},
  \citenamefont {Ryskin},\ and\ \citenamefont {Teubner}}]{Martin:1999wb}%
  \BibitemOpen
  \bibfield  {author} {\bibinfo {author} {\bibfnamefont {A.~D.}\ \bibnamefont
  {Martin}}, \bibinfo {author} {\bibfnamefont {M.}~\bibnamefont {Ryskin}}, \
  and\ \bibinfo {author} {\bibfnamefont {T.}~\bibnamefont {Teubner}},\ }\href
  {\doibase 10.1103/PhysRevD.62.014022} {\bibfield  {journal} {\bibinfo
  {journal} {Phys. Rev. D}\ }\textbf {\bibinfo {volume} {62}},\ \bibinfo
  {pages} {014022} (\bibinfo {year} {2000})},\ \Eprint
  {http://arxiv.org/abs/hep-ph/9912551} {arXiv:hep-ph/9912551} \BibitemShut
  {NoStop}%
\bibitem [{\citenamefont {Kowalski}\ \emph {et~al.}(2006)\citenamefont
  {Kowalski}, \citenamefont {Motyka},\ and\ \citenamefont
  {Watt}}]{Kowalski:2006hc}%
  \BibitemOpen
  \bibfield  {author} {\bibinfo {author} {\bibfnamefont {H.}~\bibnamefont
  {Kowalski}}, \bibinfo {author} {\bibfnamefont {L.}~\bibnamefont {Motyka}}, \
  and\ \bibinfo {author} {\bibfnamefont {G.}~\bibnamefont {Watt}},\ }\href
  {\doibase 10.1103/PhysRevD.74.074016} {\bibfield  {journal} {\bibinfo
  {journal} {Phys. Rev. D}\ }\textbf {\bibinfo {volume} {74}},\ \bibinfo
  {pages} {074016} (\bibinfo {year} {2006})},\ \Eprint
  {http://arxiv.org/abs/hep-ph/0606272} {arXiv:hep-ph/0606272} \BibitemShut
  {NoStop}%
\bibitem [{\citenamefont {Xie}\ and\ \citenamefont
  {Chen}(2018{\natexlab{a}})}]{Xie:2018ufg}%
  \BibitemOpen
  \bibfield  {author} {\bibinfo {author} {\bibfnamefont {Y.-P.}\ \bibnamefont
  {Xie}}\ and\ \bibinfo {author} {\bibfnamefont {X.}~\bibnamefont {Chen}},\
  }\href {\doibase 10.1142/S0217751X18500343} {\bibfield  {journal} {\bibinfo
  {journal} {Int. J. Mod. Phys. A}\ }\textbf {\bibinfo {volume} {33}},\
  \bibinfo {pages} {1850034} (\bibinfo {year}
  {2018}{\natexlab{a}})}\BibitemShut {NoStop}%
\bibitem [{\citenamefont {Hatta}\ \emph {et~al.}(2017)\citenamefont {Hatta},
  \citenamefont {Xiao},\ and\ \citenamefont {Yuan}}]{Hatta:2017cte}%
  \BibitemOpen
  \bibfield  {author} {\bibinfo {author} {\bibfnamefont {Y.}~\bibnamefont
  {Hatta}}, \bibinfo {author} {\bibfnamefont {B.-W.}\ \bibnamefont {Xiao}}, \
  and\ \bibinfo {author} {\bibfnamefont {F.}~\bibnamefont {Yuan}},\ }\href
  {\doibase 10.1103/PhysRevD.95.114026} {\bibfield  {journal} {\bibinfo
  {journal} {Phys. Rev. D}\ }\textbf {\bibinfo {volume} {95}},\ \bibinfo
  {pages} {114026} (\bibinfo {year} {2017})},\ \Eprint
  {http://arxiv.org/abs/1703.02085} {arXiv:1703.02085 [hep-ph]} \BibitemShut
  {NoStop}%
\bibitem [{\citenamefont {Dosch}\ \emph {et~al.}(1997)\citenamefont {Dosch},
  \citenamefont {Gousset}, \citenamefont {Kulzinger},\ and\ \citenamefont
  {Pirner}}]{Dosch:1996ss}%
  \BibitemOpen
  \bibfield  {author} {\bibinfo {author} {\bibfnamefont {H.~G.}\ \bibnamefont
  {Dosch}}, \bibinfo {author} {\bibfnamefont {T.}~\bibnamefont {Gousset}},
  \bibinfo {author} {\bibfnamefont {G.}~\bibnamefont {Kulzinger}}, \ and\
  \bibinfo {author} {\bibfnamefont {H.}~\bibnamefont {Pirner}},\ }\href
  {\doibase 10.1103/PhysRevD.55.2602} {\bibfield  {journal} {\bibinfo
  {journal} {Phys. Rev. D}\ }\textbf {\bibinfo {volume} {55}},\ \bibinfo
  {pages} {2602} (\bibinfo {year} {1997})},\ \Eprint
  {http://arxiv.org/abs/hep-ph/9608203} {arXiv:hep-ph/9608203} \BibitemShut
  {NoStop}%
\bibitem [{\citenamefont {Beuf}(2016)}]{Beuf:2016wdz}%
  \BibitemOpen
  \bibfield  {author} {\bibinfo {author} {\bibfnamefont {G.}~\bibnamefont
  {Beuf}},\ }\href {\doibase 10.1103/PhysRevD.94.054016} {\bibfield  {journal}
  {\bibinfo  {journal} {Phys. Rev. D}\ }\textbf {\bibinfo {volume} {94}},\
  \bibinfo {pages} {054016} (\bibinfo {year} {2016})},\ \Eprint
  {http://arxiv.org/abs/1606.00777} {arXiv:1606.00777 [hep-ph]} \BibitemShut
  {NoStop}%
\bibitem [{\citenamefont {H\"anninen}\ \emph {et~al.}(2018)\citenamefont
  {H\"anninen}, \citenamefont {Lappi},\ and\ \citenamefont
  {Paatelainen}}]{Hanninen:2017ddy}%
  \BibitemOpen
  \bibfield  {author} {\bibinfo {author} {\bibfnamefont {H.}~\bibnamefont
  {H\"anninen}}, \bibinfo {author} {\bibfnamefont {T.}~\bibnamefont {Lappi}}, \
  and\ \bibinfo {author} {\bibfnamefont {R.}~\bibnamefont {Paatelainen}},\
  }\href {\doibase 10.1016/j.aop.2018.04.015} {\bibfield  {journal} {\bibinfo
  {journal} {Annals Phys.}\ }\textbf {\bibinfo {volume} {393}},\ \bibinfo
  {pages} {358} (\bibinfo {year} {2018})},\ \Eprint
  {http://arxiv.org/abs/1711.08207} {arXiv:1711.08207 [hep-ph]} \BibitemShut
  {NoStop}%
\bibitem [{\citenamefont {Maris}\ and\ \citenamefont
  {Tandy}(2000{\natexlab{b}})}]{Maris:1999bh}%
  \BibitemOpen
  \bibfield  {author} {\bibinfo {author} {\bibfnamefont {P.}~\bibnamefont
  {Maris}}\ and\ \bibinfo {author} {\bibfnamefont {P.~C.}\ \bibnamefont
  {Tandy}},\ }\href {\doibase 10.1103/PhysRevC.61.045202} {\bibfield  {journal}
  {\bibinfo  {journal} {Phys. Rev. C}\ }\textbf {\bibinfo {volume} {61}},\
  \bibinfo {pages} {045202} (\bibinfo {year} {2000}{\natexlab{b}})},\ \Eprint
  {http://arxiv.org/abs/nucl-th/9910033} {arXiv:nucl-th/9910033} \BibitemShut
  {NoStop}%
\bibitem [{\citenamefont {Lappi}\ and\ \citenamefont
  {Mantysaari}(2011)}]{Lappi:2010dd}%
  \BibitemOpen
  \bibfield  {author} {\bibinfo {author} {\bibfnamefont {T.}~\bibnamefont
  {Lappi}}\ and\ \bibinfo {author} {\bibfnamefont {H.}~\bibnamefont
  {Mantysaari}},\ }\href {\doibase 10.1103/PhysRevC.83.065202} {\bibfield
  {journal} {\bibinfo  {journal} {Phys. Rev.}\ }\textbf {\bibinfo {volume}
  {C83}},\ \bibinfo {pages} {065202} (\bibinfo {year} {2011})},\ \Eprint
  {http://arxiv.org/abs/1011.1988} {arXiv:1011.1988 [hep-ph]} \BibitemShut
  {NoStop}%
\bibitem [{\citenamefont {Rezaeian}\ \emph {et~al.}(2013)\citenamefont
  {Rezaeian}, \citenamefont {Siddikov}, \citenamefont {Van~de Klundert},\ and\
  \citenamefont {Venugopalan}}]{Rezaeian:2012ji}%
  \BibitemOpen
  \bibfield  {author} {\bibinfo {author} {\bibfnamefont {A.~H.}\ \bibnamefont
  {Rezaeian}}, \bibinfo {author} {\bibfnamefont {M.}~\bibnamefont {Siddikov}},
  \bibinfo {author} {\bibfnamefont {M.}~\bibnamefont {Van~de Klundert}}, \ and\
  \bibinfo {author} {\bibfnamefont {R.}~\bibnamefont {Venugopalan}},\ }\href
  {\doibase 10.1103/PhysRevD.87.034002} {\bibfield  {journal} {\bibinfo
  {journal} {Phys. Rev.}\ }\textbf {\bibinfo {volume} {D87}},\ \bibinfo {pages}
  {034002} (\bibinfo {year} {2013})},\ \Eprint {http://arxiv.org/abs/1212.2974}
  {arXiv:1212.2974 [hep-ph]} \BibitemShut {NoStop}%
\bibitem [{\citenamefont {Xie}\ and\ \citenamefont {Chen}(2016)}]{Xie:2016ino}%
  \BibitemOpen
  \bibfield  {author} {\bibinfo {author} {\bibfnamefont {Y.-p.}\ \bibnamefont
  {Xie}}\ and\ \bibinfo {author} {\bibfnamefont {X.}~\bibnamefont {Chen}},\
  }\href {\doibase 10.1140/epjc/s10052-016-4170-1} {\bibfield  {journal}
  {\bibinfo  {journal} {Eur. Phys. J.}\ }\textbf {\bibinfo {volume} {C76}},\
  \bibinfo {pages} {316} (\bibinfo {year} {2016})},\ \Eprint
  {http://arxiv.org/abs/1602.00937} {arXiv:1602.00937 [hep-ph]} \BibitemShut
  {NoStop}%
\bibitem [{\citenamefont {Xie}\ and\ \citenamefont {Chen}(2017)}]{Xie:2017mil}%
  \BibitemOpen
  \bibfield  {author} {\bibinfo {author} {\bibfnamefont {Y.-P.}\ \bibnamefont
  {Xie}}\ and\ \bibinfo {author} {\bibfnamefont {X.}~\bibnamefont {Chen}},\
  }\href {\doibase 10.1016/j.nuclphysa.2016.12.010} {\bibfield  {journal}
  {\bibinfo  {journal} {Nucl. Phys.}\ }\textbf {\bibinfo {volume} {A959}},\
  \bibinfo {pages} {56} (\bibinfo {year} {2017})},\ \Eprint
  {http://arxiv.org/abs/1805.05901} {arXiv:1805.05901 [hep-ph]} \BibitemShut
  {NoStop}%
\bibitem [{\citenamefont {Xie}\ and\ \citenamefont
  {Chen}(2018{\natexlab{b}})}]{Xie:2018rog}%
  \BibitemOpen
  \bibfield  {author} {\bibinfo {author} {\bibfnamefont {Y.-P.}\ \bibnamefont
  {Xie}}\ and\ \bibinfo {author} {\bibfnamefont {X.}~\bibnamefont {Chen}},\
  }\href {\doibase 10.1016/j.nuclphysa.2017.12.003} {\bibfield  {journal}
  {\bibinfo  {journal} {Nucl. Phys.}\ }\textbf {\bibinfo {volume} {A970}},\
  \bibinfo {pages} {316} (\bibinfo {year} {2018}{\natexlab{b}})},\ \Eprint
  {http://arxiv.org/abs/1805.06210} {arXiv:1805.06210 [hep-ph]} \BibitemShut
  {NoStop}%
\bibitem [{\citenamefont {Golec-Biernat}\ and\ \citenamefont
  {Wusthoff}(1998)}]{GolecBiernat:1998js}%
  \BibitemOpen
  \bibfield  {author} {\bibinfo {author} {\bibfnamefont {K.~J.}\ \bibnamefont
  {Golec-Biernat}}\ and\ \bibinfo {author} {\bibfnamefont {M.}~\bibnamefont
  {Wusthoff}},\ }\href {\doibase 10.1103/PhysRevD.59.014017} {\bibfield
  {journal} {\bibinfo  {journal} {Phys. Rev.}\ }\textbf {\bibinfo {volume}
  {D59}},\ \bibinfo {pages} {014017} (\bibinfo {year} {1998})},\ \Eprint
  {http://arxiv.org/abs/hep-ph/9807513} {arXiv:hep-ph/9807513 [hep-ph]}
  \BibitemShut {NoStop}%
\bibitem [{\citenamefont {Forshaw}\ \emph {et~al.}(1999)\citenamefont
  {Forshaw}, \citenamefont {Kerley},\ and\ \citenamefont
  {Shaw}}]{Forshaw:1999uf}%
  \BibitemOpen
  \bibfield  {author} {\bibinfo {author} {\bibfnamefont {J.~R.}\ \bibnamefont
  {Forshaw}}, \bibinfo {author} {\bibfnamefont {G.}~\bibnamefont {Kerley}}, \
  and\ \bibinfo {author} {\bibfnamefont {G.}~\bibnamefont {Shaw}},\ }\href
  {\doibase 10.1103/PhysRevD.60.074012} {\bibfield  {journal} {\bibinfo
  {journal} {Phys. Rev.}\ }\textbf {\bibinfo {volume} {D60}},\ \bibinfo {pages}
  {074012} (\bibinfo {year} {1999})},\ \Eprint
  {http://arxiv.org/abs/hep-ph/9903341} {arXiv:hep-ph/9903341 [hep-ph]}
  \BibitemShut {NoStop}%
\bibitem [{\citenamefont {Iancu}\ \emph {et~al.}(2004)\citenamefont {Iancu},
  \citenamefont {Itakura},\ and\ \citenamefont {Munier}}]{Iancu:2003ge}%
  \BibitemOpen
  \bibfield  {author} {\bibinfo {author} {\bibfnamefont {E.}~\bibnamefont
  {Iancu}}, \bibinfo {author} {\bibfnamefont {K.}~\bibnamefont {Itakura}}, \
  and\ \bibinfo {author} {\bibfnamefont {S.}~\bibnamefont {Munier}},\ }\href
  {\doibase 10.1016/j.physletb.2004.02.040} {\bibfield  {journal} {\bibinfo
  {journal} {Phys. Lett. B}\ }\textbf {\bibinfo {volume} {590}},\ \bibinfo
  {pages} {199} (\bibinfo {year} {2004})},\ \Eprint
  {http://arxiv.org/abs/hep-ph/0310338} {arXiv:hep-ph/0310338} \BibitemShut
  {NoStop}%
\bibitem [{\citenamefont {Rezaeian}\ and\ \citenamefont
  {Schmidt}(2013)}]{Rezaeian:2013tka}%
  \BibitemOpen
  \bibfield  {author} {\bibinfo {author} {\bibfnamefont {A.~H.}\ \bibnamefont
  {Rezaeian}}\ and\ \bibinfo {author} {\bibfnamefont {I.}~\bibnamefont
  {Schmidt}},\ }\href {\doibase 10.1103/PhysRevD.88.074016} {\bibfield
  {journal} {\bibinfo  {journal} {Phys. Rev. D}\ }\textbf {\bibinfo {volume}
  {88}},\ \bibinfo {pages} {074016} (\bibinfo {year} {2013})},\ \Eprint
  {http://arxiv.org/abs/1307.0825} {arXiv:1307.0825 [hep-ph]} \BibitemShut
  {NoStop}%
\bibitem [{\citenamefont {Boer}\ \emph {et~al.}(2011)\citenamefont {Boer} \emph
  {et~al.}}]{Boer:2011fh}%
  \BibitemOpen
  \bibfield  {author} {\bibinfo {author} {\bibfnamefont {D.}~\bibnamefont
  {Boer}} \emph {et~al.},\ }\href@noop {} {\  (\bibinfo {year} {2011})},\
  \Eprint {http://arxiv.org/abs/1108.1713} {arXiv:1108.1713 [nucl-th]}
  \BibitemShut {NoStop}%
\bibitem [{\citenamefont {Forshaw}\ \emph {et~al.}(2004)\citenamefont
  {Forshaw}, \citenamefont {Sandapen},\ and\ \citenamefont
  {Shaw}}]{Forshaw:2003ki}%
  \BibitemOpen
  \bibfield  {author} {\bibinfo {author} {\bibfnamefont {J.~R.}\ \bibnamefont
  {Forshaw}}, \bibinfo {author} {\bibfnamefont {R.}~\bibnamefont {Sandapen}}, \
  and\ \bibinfo {author} {\bibfnamefont {G.}~\bibnamefont {Shaw}},\ }\href
  {\doibase 10.1103/PhysRevD.69.094013} {\bibfield  {journal} {\bibinfo
  {journal} {Phys. Rev. D}\ }\textbf {\bibinfo {volume} {69}},\ \bibinfo
  {pages} {094013} (\bibinfo {year} {2004})},\ \Eprint
  {http://arxiv.org/abs/hep-ph/0312172} {arXiv:hep-ph/0312172} \BibitemShut
  {NoStop}%
\bibitem [{\citenamefont {Berger}\ and\ \citenamefont
  {Stasto}(2013)}]{Berger:2012wx}%
  \BibitemOpen
  \bibfield  {author} {\bibinfo {author} {\bibfnamefont {J.}~\bibnamefont
  {Berger}}\ and\ \bibinfo {author} {\bibfnamefont {A.~M.}\ \bibnamefont
  {Stasto}},\ }\href {\doibase 10.1007/JHEP01(2013)001} {\bibfield  {journal}
  {\bibinfo  {journal} {JHEP}\ }\textbf {\bibinfo {volume} {01}},\ \bibinfo
  {pages} {001} (\bibinfo {year} {2013})},\ \Eprint
  {http://arxiv.org/abs/1205.2037} {arXiv:1205.2037 [hep-ph]} \BibitemShut
  {NoStop}%
\bibitem [{\citenamefont {Gon\c{c}alves}\ and\ \citenamefont
  {Moreira}(2020)}]{Goncalves:2020cir}%
  \BibitemOpen
  \bibfield  {author} {\bibinfo {author} {\bibfnamefont {V.~P.}\ \bibnamefont
  {Gon\c{c}alves}}\ and\ \bibinfo {author} {\bibfnamefont {B.~D.}\ \bibnamefont
  {Moreira}},\ }\href {\doibase 10.1140/epjc/s10052-020-8043-2} {\bibfield
  {journal} {\bibinfo  {journal} {Eur. Phys. J. C}\ }\textbf {\bibinfo {volume}
  {80}},\ \bibinfo {pages} {492} (\bibinfo {year} {2020})},\ \Eprint
  {http://arxiv.org/abs/2003.11438} {arXiv:2003.11438 [hep-ph]} \BibitemShut
  {NoStop}%
\bibitem [{\citenamefont {Agostini}\ \emph {et~al.}(2020)\citenamefont
  {Agostini} \emph {et~al.}}]{Agostini:2020fmq}%
  \BibitemOpen
  \bibfield  {author} {\bibinfo {author} {\bibfnamefont {P.}~\bibnamefont
  {Agostini}} \emph {et~al.} (\bibinfo {collaboration} {LHeC, FCC-he Study
  Group}),\ }\href@noop {} {\  (\bibinfo {year} {2020})},\ \Eprint
  {http://arxiv.org/abs/2007.14491} {arXiv:2007.14491 [hep-ex]} \BibitemShut
  {NoStop}%
\bibitem [{\citenamefont {Anderle}\ \emph {et~al.}(2021)\citenamefont {Anderle}
  \emph {et~al.}}]{Anderle:2021wcy}%
  \BibitemOpen
  \bibfield  {author} {\bibinfo {author} {\bibfnamefont {D.~P.}\ \bibnamefont
  {Anderle}} \emph {et~al.},\ }\href@noop {} {\  (\bibinfo {year} {2021})},\
  \Eprint {http://arxiv.org/abs/2102.09222} {arXiv:2102.09222 [nucl-ex]}
  \BibitemShut {NoStop}%
\end{thebibliography}%

\end{document}